\documentclass[11pt,english]{article}
\usepackage[T1]{fontenc}
\usepackage[utf8]{inputenc}
\usepackage[a4paper]{geometry}
\usepackage{color}
\usepackage{enumitem}
\usepackage{bm}
\usepackage{amsmath}
\usepackage{amsthm}
\usepackage{amssymb}
\usepackage{graphicx}
\usepackage{setspace}
\usepackage[authoryear]{natbib}
\PassOptionsToPackage{normalem}{ulem}
\usepackage{ulem}
\usepackage{cases}
\usepackage[title]{appendix}
\usepackage{cancel}

\makeatletter

\numberwithin{table}{section}
\numberwithin{figure}{section}
\theoremstyle{plain}
\newtheorem{thm}{\protect\theoremname}[section]
\newtheorem{rem}[thm]{\protect\remarkname}
\theoremstyle{plain}
\newtheorem{assumption}[thm]{\protect\assumptionname}
\theoremstyle{definition}
\newtheorem{defn}[thm]{\protect\definitionname}
\theoremstyle{plain}
\newtheorem{lem}[thm]{\protect\lemmaname}
\theoremstyle{plain}
\newtheorem{prop}[thm]{\protect\propositionname}
\theoremstyle{definition}
\newtheorem{example}[thm]{\protect\examplename}

\usepackage{amsfonts,amssymb,amsthm,amsbsy,amsmath,mathrsfs}
\usepackage{graphicx,rotating,array,sectsty}
\usepackage{natbib}
\usepackage[linkcolor=blue,citecolor=blue, colorlinks]{hyperref}
\usepackage{cases}
\usepackage[overload]{empheq}

\newcommand{\mcl}[1]{\mathcal{#1}}

\newcommand{\mbb}[1]{\mathbb{#1}}

\newcommand{\dt}{\mathrm{d}}

\newcommand{\esssup}{\mathop{\mathrm{esssup}}}

\def\eed{{\hbox{ }\hfill {$\lozenge$}}}

\def\Pbb{\mathbb{P}}

\def\Rbb{\mathbb{R}}

\def\Ebf{\mathbf{E}}

\def\Ac{\mathcal{A}}

\def\Dc{\mathcal{D}}
\def\Ec{\mathcal{E}}
\def\Fc{\mathcal{F}}
\def\Gc{\mathcal{G}}

\def\Qc{\mathcal{Q}}

\def\Tc{\mathcal{T}}

\def\Qsc{\mathscr{Q}}

\def\Ebf{\mathbf{E}}

\def\Rbb{\mathbb{R}}

\def\trieq{\triangleq}

\DeclareMathOperator*{\maxi}{maximize\,}

\newcommand{\overbar}[1]{\mkern 1.5mu\overline{\mkern-1.5mu#1\mkern-1.5mu}\mkern 1.5mu}
\newcommand{\overtilde}[1]{\mkern 1.5mu\widetilde{\mkern-1.5mu#1\mkern-1.5mu}\mkern 1.5mu}
\newcommand{\overhat}[1]{\mkern 1.5mu\widehat{\mkern-1.5mu#1\mkern-1.5mu}\mkern 1.5mu}

\setcitestyle{round}

\numberwithin{equation}{section}

\allowdisplaybreaks[4]

\ifdefined\showcaptionsetup
 \PassOptionsToPackage{caption=false}{subfig}
\fi
\usepackage{subfig}
\makeatother

\usepackage{babel}
\providecommand{\assumptionname}{Assumption}
\providecommand{\definitionname}{Definition}
\providecommand{\examplename}{Example}
\providecommand{\lemmaname}{Lemma}
\providecommand{\propositionname}{Proposition}
\providecommand{\remarkname}{Remark}
\providecommand{\theoremname}{Theorem}

\begin{document}
\title{Time-Consistent Portfolio Selection for Rank-Dependent Utilities in
an Incomplete Market\footnote{This research is supported by the National Key R\&D Program of China (NO. 2020YFA0712700) and the National Natural Science Foundation of China (NOs. 12431017, 12471447, 12071146, 11971301).}}

\author{Jiaqin Wei\thanks{Key Laboratory of Advanced Theory and Application in Statistics and Data Science-MOE, School of Statistics, East China Normal University, Shanghai 200062, China. Email: jqwei@stat.ecnu.edu.cn.} \and  Jianming Xia\thanks{Key Laboratory of Random Complex Structures and Data Science (RCSDS), National Center for Mathematics and Interdisciplinary Sciences (NCMIS), Academy of Mathematics and Systems Science, Chinese Academy of Sciences, Beijing 100190, China. Email: xia@amss.ac.cn.} \and  Qian Zhao\thanks{School of Statistics and Information, Shanghai University of International Business and Economics, Shanghai 201620, China. E-mail: qzhao31@163.com.}}

\maketitle
\begin{abstract}
We investigate the portfolio selection problem for an agent with rank-dependent utility in an incomplete financial market. For a constant-coefficient market and CRRA utilities, we characterize the deterministic strict equilibrium strategies. In the case of  time-invariant probability weighting function, we provide a comprehensive characterization of the deterministic strict equilibrium strategy. The unique non-zero equilibrium, if exists, can be determined by solving an autonomous ODE.
In the case of time-variant probability weighting functions, we observe that there may be infinitely many non-zero deterministic strict equilibrium strategies, which are derived from the positive solutions to a nonlinear singular ODE. By specifying the maximal solution to the singular ODE, we are able to identify all the positive solutions. In addition, we address the issue of selecting an optimal strategy from the numerous equilibrium strategies available.

\vspace{3mm}
\end{abstract}

\section{Introduction}

It is well documented that the expected utility (EU) theory is unable
to explain many decision-making behaviors, such as the famous Allais paradox (\citet*{a1953}).
This motivates emergence of various non-EU theories, including the rank-dependent utility (RDU) theory of \citet*{q1982}.  A RDU consists of an (outcome) utility function and a probability weighting
function. A probability weighting function helps the RDU theory explain
some of the paradoxes encountered by the EU theory.

Because of its superiority over EU, RDU has attracted much attention
from researchers. In particular, the problem of the continuous-time RDU maximization has been intensively investigated by 
\citet*{cd2006, cd2011}, \citet*{JinZhou2008} (where they consider the cumulative prospect theory, which includes the
RDU theory as a special case),   \citet*{XiaZhou2016}, and \citet*{Xu2016}, among others.
Du to the presence of the probability weighting function, the optimal strategy is, however, usually
time-inconsistent: the current optimal strategy is not necessarily optimal in the future. 

\citet*{hjz2021} is the first work to address the issue of time-inconsistency in continuous-time portfolio selection with RDU: they derive the intra-person equilibrium strategies. The idea
of intra-person equilibrium can go back to \citet*{s55}, where 
the time-inconsistent problem is regarded as a non-cooperative
game among the agent’s selves at different times and the corresponding sub-game perfect Nash equilibria are suggested as the solution.
Following Strotz's idea, there is a vast literature on
the intra-person equilibria for various time-inconsistent optimization
problems. See \citet*{Ekeland2006} (published version: \citet*{Ekeland2010}), \citet*{ep08}, \citet*{bc10}, 
\citet*{hjz12,hjz17}, \citet*{y122,y17}, \citet*{bmz14}, \citet*{bkm17}, \citet*{ww17},
and \citet*{hj2021}, among others.

The probability weighting function makes the RDU a highly nonlinear functional of the terminal wealth or its distribution. 
It is very difficult to investigate the intra-person equilibrium with a RDU agent for a general market model. So \citet*{hjz2021} consider a \emph{complete} market with deterministic coefficients. For a general concave utility function $U:\mathbb{R}\to\mathbb{R}$, they make an ansatz that the marginal utility $U'(X_T)$ of the terminal wealth $X_T$ of an equilibrium strategy is proportional to a \emph{revised} state-price density $\overbar{\rho}(T)$, which is determined by multiplying the market price of risk function $\theta(\cdot)$ by a scaling function $\lambda(\cdot)$:
$$\overbar\rho(t)=\exp\left(-{\frac{1}{2}}\int_0^t|\lambda(s)\theta(s)|^2\dt s-\int_0^t\lambda(s)\theta(s)^\top\dt W(s)\right).$$   
Then they derive a highly nonlinear ordinary differential equation (ODE) for the scaling function $\lambda(\cdot)$, which is singular at the terminal time $T$, and show that an equilibrium can be generated by a \emph{positive} solution to the ODE,  the existence  of which is also established. It is well known that, in the EU case, the marginal utility of the optimal terminal wealth is proportion to the state-price density.  
Therefore, their approach can be regraded as a nontrivial extension of the martingale/dual approach from the EU case to the RDU case. 

This paper considers the dynamic portfolio selection problem with a RDU agent in an \emph{incomplete}
financial market with constant coefficients (the extension to the deterministic coefficient case is easy). 
In contrast to the existing literature, which usually investigates the so-called \emph{weak} equilibria, this paper investigates the \emph{strict} equilibria; see Definition \ref{def:eq}. 
To highlight the effect of the probability weighting function, we focus on the constant-relative-risk-aversion (CRRA) utility functions. The probability weighting function makes it very difficult (if not impossible) to apply the approach of \citet*{hjz2021} to an incomplete market. The homotheticity of the CRRA utility, however, makes us consider the deterministic strict equilibrium strategies (DSESes) and then we are able to directly derive the equation for the equilibrium strategies without applying the revised martingale/dual approach of \citet*{hjz2021}. 

This paper essentially consists of two parts.   

In the first part (Section \ref{sec:dtmn}), the probability weighting function does not change over time. Under a mild technical assumption (Assumption \ref{assu:DistFun}), we make a thorough discussion that covers all possible cases.  We identify the cases where there is no DSES and the cases where $\bm{0}$ is the unique DSES. For the other cases, we show that there is a unique non-zero DSES, which is worked out in closed form by solving an autonomous ODE.  In particular, we allow for convex CRRA utility functions,  whereas the utility function is concave in \citet*{hjz2021} as well as in the aforementioned literature on the continuous-time RDU maximization. Moreover, we do not restrict the shape of the probability weighting function, whereas \citet*{hjz2021} assume that the probability weighting function is either convex
or inverse-S shaped to guarantee the validity of their inequality (8). 

In the second part (Section \ref{sec:td}), the probability weighting functions change over time and their shape is not restricted. For technical reasons, we consider those CRRA utilities that is more risk averse than or equal to the logarithmic utility. In this case,  non-zero DSESes can be generated from the positive solutions of a nonlinear backward ODE, which is singular at the terminal time. Under various mild technical assumptions, we establish the existence as well as the uniqueness of the positive solutions to the backward ODE. The establishment of the existence is similar to \citet*{hjz2021}, but our condition is weaker. More importantly, we get some deeper discussions as follows. 
\begin{itemize}
\item In general, there are infinitely many positive solutions and so it is difficult (if not impossible) to find all positive solutions by solving the  ODE backwardly. We can approximate the maximal positive solution by the solution to the backward ODE with the terminal value being positive but sufficiently small. However, we can not find out the other positive solutions in this way. This motives us to look for a ``forward” method to get all positive solutions. The forward method is implementable because the ODE is regular at any time but the terminal time. 
\item Every positive solution generates a DSES. A question then arises: which one should the agent choose from these infinitely many equilibrium strategies? 
We suggest that the agent should maximize the RDU of the initial time over all DSESes. This leads to the discussion on the so-called \emph{optima} equilibria. 
To our best knowledge, there is no existing work on the optimal equilibria for the time-inconsistent \emph{control} problems but three recent papers: \citet*{hz2019,hz2020mf} and \citet*{hw2021}, where the optimal equilibria for time-inconsistent \emph{stopping} problems are discussed. 
\end{itemize}

Another literature related to this work is \citet*{hsz2021}, which studies a portfolio selection problem under a forward rank-dependent performance criterion. The authors first construct time-dependent probability weighting functions and outcome utility functions such that the portfolio selection problem is time-consistent and then characterize the optimal wealth process by investigating an equivalent problem under EU theory. This paper, however, in line with \citet*{hjz2021}, focuses on characterizing equilibrium strategies for a given time-inconsistent portfolio selection problem under RDU.

The rest of the paper is organized as follows. Section 2 formulates
the investment problem for a RUD agent.
Section 3 makes a thorough analysis when the probability weighting function is time-invariant.
Section 4 discusses the case of time-variant
probability weighting functions. Section 5  concludes the paper. The appendices investigate the ODEs related to the equilibria.

\section{Problem Formulation \protect\label{sec:formulation}}

\subsection{Financial Market and Trading Strategies}

Let $(\Omega,\Fc,\{\Fc_{t}\}_{t\in[0,T]},\mathbb{P})$ be a filtered
complete probability space on which a standard $d$-dimensional Brownian
motion $\boldsymbol{W}=(W_{1},W_{2},\cdots,W_{d})^{\top}$ (the
superscript $\top$ denotes the transposition)
is defined, where the constant $T>0$ is a fixed time horizon and the filtration
$\{\Fc_{t}\}_{t\in[0,T]}$ 
is the $\mathbb{P}$-augmentation
of the natural filtration of $\boldsymbol{W}$.

We consider a financial market consisting of $n+1$ assets:
one bond (risk-free asset) and $n$ stocks (risky assets), where $n\le d$
is a positive integer.
The price $S_{0}$ of the risk-free bond satisfies 
\begin{equation*}
\dt S_{0}(s)=rS_{0}(s)\dt s,\quad s\in[0,T],
\end{equation*}
where constant $r$ is the risk-free return rate. 
For $i=1,2,\cdots,n,$ the price $S_{i}$ of the $i$-th stock is
given by 
\begin{equation*}
\dt S_{i}(s)=S_{i}(s)\left[(r+\mu_{i})\dt s+\sum_{j=1}^{d}\sigma_{ij}\dt W_{j}(s)\right],\quad s\in[0,T],
\end{equation*}
where constant $\mu_{i}$ is the  expected excess return rate of the $i$-th stock and constants $\sigma_{ij}$, $j=1,2,\cdots,d$, are the corresponding volatility rates. 

Let $\bm{\mu}\trieq(\mu_{1},\cdots,\mu_{n})^{\top}$ and $\bm{\sigma}\trieq(\sigma_{ij})_{1\leq i\leq n,1\leq j\leq d}$. In this paper, we always assume that $\bm{\sigma}\bm{\sigma}^{\top}$ is positive definite.
Obviously, the financial market is incomplete if $n<d$ and is complete if $n=d$.
For brevity, we only consider the constant coefficient model of $(r,\bm{\mu},\bm{\sigma})$ in this paper. The extension to the deterministic coefficient model is easy.
 
An agent trades the bond and the stocks continuously within the time
horizon $[0,T]$. A trading strategy is an $n$-dimensional progressive process $\boldsymbol{\pi}=(\pi_{1},\pi_{2},\cdots,\pi_{n})^{\top}$
such that $\int_0^T\left|\boldsymbol{\pi}^\top(t)\bm{\mu}\right|\dt t +\int_{0}^{T}\left|\boldsymbol{\pi}^\top(t)\bm{\sigma}\right|^{2}\dt t<\infty$
a.s. ($|\bm{v}|=\sqrt{\bm{v}^\top\bm{v}}$ for a vector $\bm{v}$). Here, each $\pi_{i}(t)$ is the proportion of the wealth invested
in the $i$-th stock at time $t\in[0,T]$. The self-financing wealth
process $X^{\boldsymbol{\pi}}$ evolves as 
\begin{equation*}
\frac{\dt X^{\boldsymbol{\pi}}(s)}{X^{\boldsymbol{\pi}}(s)}=\left[r+\boldsymbol{\pi}^{\top}(s)\bm{\mu}\right]\dt s+\bm{\pi}^{\top}(s)\bm{\sigma}\dt\bm{W}(s),\quad s\in[0,T].
\end{equation*}
Given the wealth $x$ at time $t$, the terminal wealth is 
\begin{equation*}
X^{\boldsymbol{\pi}}(T)=x\exp\left\{ \int_{t}^{T}\left(r+\boldsymbol{\pi}^{\top}(s)\bm{\mu}-\frac{1}{2}\left|\bm{\pi}^{\top}(s)\bm{\sigma}\right|^{2}\right)\dt s+\int_{t}^{T}\bm{\pi}^{\top}(s)\bm{\sigma}\dt\bm{W}(s)\right\}. 
\end{equation*}

\subsection{Rank-Dependent Utility}\label{sec:rdu}

The regular conditional probability measure w.r.t. $\Fc_{t}$ is denoted
by $\mathbb{P}_{t}$. For every random variable $X$, its distribution
function, denoted by $F_{X}$, is increasing\footnote{Herein, ``increasing'' means ``non-decreasing'' and ``decreasing''
means ``non-increasing''.} and right-continuous. Its conditional distribution function w.r.t.
$\Fc_{t}$ is denoted by $F_{X}^{t}$. For every random variable
$X$, its quantile function $Q_{X}:(0,1)\to\Rbb$ is defined by 
\[
Q_{X}(p)\trieq\inf\{x\in\Rbb\,|\,F_{X}(x)>p\},\quad p\in(0,1).
\]
Its conditional quantile function $Q_{X}^{t}$ w.r.t. $\Fc_{t}$
is similarly defined. For more details about quantile functions, see
\citet*[Appendix A.3]{FS2016}. We emphasize that the superscript $t$
of $F_{X}^{t}$ and $Q_{X}^{t}$ denotes the dependence
on $t$ rather than the power of $t$.

The agent's time-$t$ preference over terminal wealth
$X$ is represented by 
\begin{equation*}
\Ebf_{t}^{w}[U(X)]\trieq\int U(X)\,\dt(w\circ\Pbb_{t}),
\end{equation*}
where 
\[
\int U(X)\,\dt(w\circ\Pbb_{t})\triangleq\int U(x)\,\dt\overbar{w}\left(F_{X}^{t}(x)\right)
\]
is the RDU with (outcome) utility function $U$
for terminal wealth $X$ and probability weighting function (probability distortion function) $w$.
In the above (and hereafter) $\overbar{w}$ denotes the dual of
$w$, given by 
\[
\overbar{w}(p)\triangleq1-w(1-p),\quad p\in[0,1].
\]
Note that if $w$ is continuously differentiable, then 
\[
\int U(X)\,\dt(w\circ\Pbb_{t})=\int U(x)w^{\prime}\left(1-F_{X}^{t}(x)\right)\,\dt F_{X}^{t}(x).
\]
Hence, we have here an additional term $w^{\prime}(1-F_{X}^{t}(x))$
serving as the weight on every wealth level $x$ when calculating
the expected utility. The weight depends on the rank, $1-F_{X}^{t}(x)$,
of level $x$ over all possible realizations of $X$. For more details
about RDU, see the textbook by \citet*{Wakker2010}.

In terms of conditional quantile functions, the RDU $\Ebf_{t}^{w}[U(X)]$
can be written as 
\begin{equation*}
\Ebf_{t}^{w}[U(X)]=\int_{0}^{1}U\left(Q_{X}^{t}(p)\right)\dt\overbar{w}(p).
\end{equation*}

Following \citet*{hjz2021}, we impose the following assumption on
the probability weighting function $w$. 
\begin{assumption}
\label{assu:DistFun} $w:[0,1]\to[0,1]$ is strictly increasing and
continuously differentiable, $w(0)=0$ and $w(1)=1$. There exist
some $c>0$ and $\alpha\in(-1,0)$ such that 
\[
w'(p)\leq c\left[p^{\alpha}+(1-p)^{\alpha}\right]\quad\forall p\in(0,1).
\]
\end{assumption}

\begin{defn}
A strategy $\boldsymbol{\pi}$ is called \emph{admissible}
if $\int_{0}^{1}\left|U\left(Q_{X^{\boldsymbol{\pi}}(T)}^{t}(p)\right)\right|\dt\overbar{w}(p)<\infty$
for every $t\in[0,T)$. Denoted by $\Ac$ the set of all admissible
strategies. 
\end{defn}

For notational simplicity, let 
\[
J(t,x;\boldsymbol{\pi})\trieq\Ebf_{t}^{w}[U(X^{\boldsymbol{\pi}}(T))],\quad\boldsymbol{\pi}\in\Ac,\,t\in[0,T),\,x\in(0,\infty),
\]
where $x$ is the initial wealth at time $t$. It is well known that
the problem of maximizing $J(t,x;\boldsymbol{\pi})$ is time-inconsistent
(see \citet*{hjz2021}). In this paper, we aim to find out the so-called
equilibrium strategies.

For a trading strategy $\boldsymbol{\pi}$, $t\in[0,T)$, $\varepsilon\in[0,T-t)$
and $\boldsymbol{\kappa}\in L^{\infty}(\Fc_{t};\mathbb{R}^{n})$,
let $\boldsymbol{\pi}^{t,\varepsilon,\boldsymbol{\kappa}}$
be a strategy given by 
\begin{equation*}
\boldsymbol{\pi}^{t,\varepsilon,\boldsymbol{\kappa}}(s)=\begin{cases}
\boldsymbol{\pi}(s)+\boldsymbol{\kappa}, & s\in[t,t+\varepsilon),\\
\boldsymbol{\pi}(s), & \text{otherwise},
\end{cases}
\end{equation*}
where $L^{\infty}(\Fc_{t};\mathbb{R}^{n})$ is the set of all bounded,
$\mcl F_{t}$-measurable, $n$-dimensional  random vectors.
Strategy $\boldsymbol{\pi}^{t,\varepsilon,\boldsymbol{\kappa}}$
serves as a perturbation of $\boldsymbol{\pi}$. For notational
simplicity, we write $X^{t,\varepsilon\boldsymbol{,\kappa}}$ for
$X^{\boldsymbol{\pi}^{t,\varepsilon,\boldsymbol{\kappa}}}$. Obviously, $\bm{\pi}=\bm{\pi}^{t,0,\bm{\kappa}}$.

\begin{defn}
\label{def:eq} An admissible strategy $\boldsymbol{\pi}$ is
called a \emph{strict equilibrium strategy} (SES)
if, for every $t\in[0,T)$ and
for every $\boldsymbol{\kappa}\in L^{\infty}(\Fc_{t};\mathbb{R}^{n})\setminus\{\boldsymbol{0}\}$
such that  $\boldsymbol{\pi}^{t,\varepsilon,\boldsymbol{\kappa}}\in\Ac$
for all sufficiently small $\varepsilon>0$, it holds that \begin{equation}
\mathop{\lim}\limits_{\varepsilon\downarrow0}\esssup_{\varepsilon_{0}\in(0,\varepsilon)}\frac{J(t,x;\boldsymbol{\pi}^{t,\varepsilon_{0},\boldsymbol{\kappa}})-J(t,x;\boldsymbol{\pi})}{\varepsilon_{0}}<0\quad\text{a.s. }\forall x\in(0,\infty),\label{eq:eqs:strict}
\end{equation}
where $\varepsilon\downarrow0$ stands for that $\varepsilon>0$ and $\varepsilon$ converges to $0$. 
If $\bm{\pi}$ only satisfies a condition weaker than \eqref{eq:eqs:strict}: 
\begin{equation*}
\mathop{\lim}\limits_{\varepsilon\downarrow0}\esssup_{\varepsilon_{0}\in(0,\varepsilon)}\frac{J(t,x;\boldsymbol{\pi}^{t,\varepsilon_{0},\boldsymbol{\kappa}})-J(t,x;\boldsymbol{\pi})}{\varepsilon_{0}}\le0\quad\text{a.s. }\forall x\in(0,\infty),
\end{equation*}
then $\bm{\pi}$ is called a \emph{weak equilibrium strategy}.
\end{defn}

\begin{rem} Obviously, every SES is a weak equilibrium strategy. Condition \eqref{eq:eqs:strict} implies that $J(t,x;\boldsymbol{\pi}^{t,\varepsilon,\boldsymbol{\kappa}})<J(t,x;\boldsymbol{\pi})$ a.s. for every sufficiently small $\varepsilon>0$, similar conditions of which are usually used to define the so-called strong or regular equilibrium strategies; see \citet*{hj2021} for example.
This paper investigates the SESes. The weak equilibrium strategies,  which the existing literature usually concerns, can be investigated similarly. 
\end{rem}

In this paper we focus on the case of CRRA utility to highlight the effect of probability weighting function. Hereafter, we always assume
that the utility function $U$ is given by 
\begin{equation}
U(x)=\begin{cases}
\frac{1}{\gamma}x^{\gamma},\quad & \gamma\ne0,\\
\log x,\quad & \gamma=0,
\end{cases}\qquad x>0,\label{eq:Upower}
\end{equation}
where $\gamma\in\Rbb$. In the literature it is usually assume that
$\gamma<1$, implying $U$ is risk averse. In this paper, however,
the other case $\gamma\ge 1$ is not ruled out. 
In particular, when $\gamma=1$, the RDU reduces to the dual utility of \citet*{y1987}. We still call $U$ a CRRA utility function when $\gamma>1$. In this case, the relative risk aversion index is negative. 

\begin{rem}
We will see that the pre-commitment solution to the RDU maximization problem does not exist, under mild conditions, when $\gamma\ge 1$ and $n=d$.  Suppose $n=d$. In this case, the market is complete and there is a unique pricing kernel $\rho$, which is log-normally distributed. It is well known that the quantile formulation of the RDU maximization is
\begin{equation*}
\begin{split}
\maxi_{Q\in\Qsc}\quad &{\frac{1}{\gamma}}\int_0^1 (Q(p))^\gamma\dt \overbar{w}(p)\\
\text{subject to}\quad&\int_0^1 Q(p)Q_\rho(1-p)\dt p\le x,
\end{split}
\end{equation*} 
where $x>0$ is the initial wealth and 
$$\Qc=\{Q:(0,1)\to(0,\infty) \text{ is increasing and right continuous}\};$$
c.f. \citet*{cd2006} and \citet*{JinZhou2008} among others. 
For every $j\ge1$, consider 
\[
Q_{j}(p)=\begin{cases}
c_{j}, & p\in[p_{j},1),\\
\varepsilon_{j}, & p\in(0,p_{j}),
\end{cases}
\]
where $\varepsilon_{j}\in\left(0,\frac{x}{\Ebf[\rho]}\right)$, $\varepsilon_{j}\downarrow0$, $p_{j}\in(0,1)$, $p_{j}\uparrow1$
and 
\[
c_{j}=\frac{x-\varepsilon_{j}\int_{0}^{p_{j}}Q_{\rho}(1-p)\dt p}{\int_{p_{j}}^{1}Q_{\rho}(1-p)\dt p}\geq\frac{x-\varepsilon_{j}\Ebf[\rho]}{\int_{p_{j}}^{1}Q_{\rho}(1-p)\dt p}.
\]
It is easy to verify that
\[
\int_{0}^{1}Q_{j}(p)Q_{\rho}(1-p)\dt p=x. 
\]
Let $\gamma\ge1$. Then
\begin{align*}
\frac{1}{\gamma}\int_{0}^{1}\left(Q_{j}(p)\right)^{\gamma}\dt\overbar{w}(p)  = & \frac{1}{\gamma}\left(\int_{0}^{p_{j}}\varepsilon_{j}^{\gamma}\dt\overbar{w}(p)+\int_{p_{j}}^{1}c_{j}^{\gamma}\dt\overbar{w}(p)\right)\\
  = & \frac{1}{\gamma}\left[\varepsilon_{j}^{\gamma}\overbar{w}(p_{j})+c_{j}^{\gamma}\left(1-\overbar{w}(p_{j})\right)\right]\\
  \geq & \frac{1}{\gamma}c_{j}^{\gamma}w(1-p_{j})\\
  \geq & \frac{1}{\gamma}\left(x-\varepsilon_{j}\Ebf[\rho]\right)^{\gamma}\frac{w(1-p_{j})}{\left(\int_{p_{j}}^{1}Q_{\rho}(1-p)\dt p\right)^{\gamma}}.
\end{align*}
Therefore,
\begin{align*}
\liminf_{j\rightarrow\infty}\frac{1}{\gamma}\int_{0}^{1}
\left(Q_{j}(p)\right)^{\gamma}\dt\overbar{w}(p) 
 \geq & \frac{x^{\gamma}}{\gamma}\liminf_{j\rightarrow\infty}
 \left(\frac{w(1-p_{j})}{\int_{p_{j}}^{1}Q_{\rho}(1-p)\dt p}\right)^{\gamma}(w(1-p_{j}))^{1-\gamma}\\
 = & \frac{x^{\gamma}}{\gamma}\liminf_{j\rightarrow\infty}
 \left(\frac{w'(1-p_{j})}{Q_{\rho}(1-p_{j})}\right)^{\gamma}
 (w(1-p_{j}))^{1-\gamma}\\
  = & \infty
\end{align*}
provided either of the following conditions is satisfied: (i) $\gamma>1, \lim_{p\rightarrow0}\frac{w'(p)}{Q_{\rho}(p)}>0$;
(ii) $\gamma=1, \lim_{p\rightarrow0}\frac{w'(p)}{Q_{\rho}(p)}=\infty$. Consequently, under either of the above conditions (i) and (ii), the pre-commitment solution to the RDU maximization problem does not exist.
\end{rem}

\section{Deterministic Strict Equilibrium Strategies}\label{sec:dtmn}

Denote by $\Dc$ all the deterministic and right-continuous functions
$\boldsymbol{\pi}:[0,T)\to\Rbb^{n}$ such that 
\[
\int_{0}^{T}\left|\bm{\pi}^{\top}(s)\bm{\mu}\right|\dt s+
\int_{0}^{T}\left|\bm{\pi}^{\top}(s)\bm{\sigma}\right|^{2}\dt s<\infty.
\]
We say that $\boldsymbol{\pi}$
is a \emph{deterministic strict equilibrium strategy} (DSES) if $\boldsymbol{\pi}\in\Dc$
is  an SES.
Since the CRRA utility functions are homothetic and the coefficients of the market model are constant, it is natural to look for  the equilibrium strategies in $\Dc$.  

In the following, we consider a fixed $\boldsymbol{\pi}\in\Dc$. For any $t\in[0,T)$, $\varepsilon\in[0,T-t)$ and $\boldsymbol{\kappa}\in L^{\infty}(\Fc_{t};\Rbb^{n})$, 
given the initial wealth $X_{t}^{\boldsymbol{\pi}}=x$,
we have
\begin{align*}
X^{t,\varepsilon,\boldsymbol{\kappa}}(T)  = & x\exp\left\{ r(T-t)+\int_{t}^{t+\varepsilon}
\left(\boldsymbol{\pi}(s)+\boldsymbol{\kappa}\right)^\top\bm{\mu}\dt s+\int_{t+\varepsilon}^{T}\boldsymbol{\pi}^{\top}(s)\bm{\mu}\dt s\right.\\
  & \left.-\frac{1}{2}\left[\int_{t}^{t+\varepsilon}\left|\left(\boldsymbol{\pi}(s)+\boldsymbol{\kappa}\right)^{\top}\bm{\sigma}\right|^{2}\dt s+\int_{t+\varepsilon}^{T}\left|\bm{\pi}^{\top}(s)\bm{\sigma}\right|^{2}\dt s\right]\right.\\
  & \left.+\left[\int_{t}^{t+\varepsilon}\left(\boldsymbol{\pi}(s)+\boldsymbol{\kappa}\right)^{\top}\bm{\sigma}\dt\bm{W}(s)+\int_{t+\varepsilon}^{T}\bm{\pi}^{\top}(s)\bm{\sigma}\dt\bm{W}(s)\right]\right\} .
\end{align*}
Then the distribution of $\ln X^{t,\varepsilon,\boldsymbol{\kappa}}(T)$
conditioned on $\mathcal{F}_{t}$ is 
\[
N\left(\ln x+r(T-t)+g(\varepsilon;t,\boldsymbol{\kappa})-\frac{1}{2}H(\varepsilon;t,\boldsymbol{\kappa}),H(\varepsilon;t,\boldsymbol{\kappa})\right),
\]
where 
\begin{align*}
g(\varepsilon;t,\boldsymbol{\kappa})  \triangleq & \int_{t}^{T}\boldsymbol{\pi}^{\top}(s)\bm{\mu}\dt s+\boldsymbol{\kappa}^\top\bm{\mu}\varepsilon,\\
H(\varepsilon;t,\boldsymbol{\kappa})  \triangleq & \int_{t}^{t+\varepsilon}\left|\left(\boldsymbol{\pi}(s)+\boldsymbol{\kappa}\right)^{\top}\bm{\sigma}\right|^{2}\dt s+\int_{t+\varepsilon}^{T}\left|\bm{\pi}^{\top}(s)\bm{\sigma}\right|^{2}\dt s.
\end{align*}
Therefore 
\[
Q_{X^{t,\varepsilon,\boldsymbol{\kappa}}(T)}^{t}(p)=x\exp\left\{ r(T-t)+g(\varepsilon;t,\boldsymbol{\kappa})-\frac{1}{2}H(\varepsilon;t,\boldsymbol{\kappa})+\sqrt{H(\varepsilon;t,\boldsymbol{\kappa})}Q_{\xi}(p)\right\} ,\quad p\in(0,1).
\]
Hereafter, $\xi$ always denotes a standard normal random variable: $\xi\sim N(0,1)$.

Obviously, $\Dc\subset\Ac$ and  $\boldsymbol{\pi}^{t,\varepsilon,\boldsymbol{\kappa}}\in\Ac$
for all $t\in[0,T)$, $\varepsilon\in(0,T-t)$ and $\boldsymbol{\kappa}\in L^{\infty}(\Fc_{t};\Rbb^{n})$.
Therefore,  $\boldsymbol{\pi}$ is an SES if and only
if, for every $t\in[0,T)$, 
\begin{equation}
\mathop{\lim}\limits_{\varepsilon\downarrow0}\esssup_{\varepsilon_{0}\in(0,\varepsilon)}\frac{J(t,x;\boldsymbol{\pi}^{t,\varepsilon_{0},\boldsymbol{\kappa}})-J(t,x;\boldsymbol{\pi})}{\varepsilon_{0}}<0\text{ a.s.}\quad\forall\,\boldsymbol{\kappa}\in L^{\infty}(\Fc_{t};\Rbb^{n})\setminus\{\boldsymbol{0}\}\text{ and }x\in(0,\infty).\label{eq:mv01:strict}
\end{equation}

When $\gamma\ne0$, the RDU of the terminal wealth $X^{t,\varepsilon,\boldsymbol{\kappa}}(T)$
at time $t$ is 
\begin{align*}
f(\varepsilon;t,\boldsymbol{\kappa})\trieq & J(t,x;\boldsymbol{\pi}^{t,\varepsilon_{0},\boldsymbol{\kappa}})\\ 
= & \int_{0}^{1}\frac{1}{\gamma}x^{\gamma}\exp\left\{ \gamma\left[r(T-t)+g(\varepsilon;t,\boldsymbol{\kappa})-\frac{1}{2}H(\varepsilon;t,\boldsymbol{\kappa})+\sqrt{H(\varepsilon;t,\boldsymbol{\kappa})}Q_{\xi}(p)\right]\right\} \dt\overbar{w}(p)\\
= & \frac{1}{\gamma}x^{\gamma}\exp\left\{ \gamma\left[r(T-t)+g(\varepsilon;t,\boldsymbol{\kappa})-\frac{1}{2}H(\varepsilon;t,\boldsymbol{\kappa})\right]\right\} h\left(-\gamma\sqrt{H(\varepsilon;t,\boldsymbol{\kappa})}\right),
\end{align*}
where the function $h:\Rbb\to(0,\infty)$ is defined by 
\begin{equation}
h(x)\trieq\int_{0}^{1}e^{-xQ_{\xi}(p)}\dt\overbar{w}(p), 
\quad x\in\mathbb{R}.\label{eq:h}
\end{equation}
The function $h$ defined by \eqref{eq:h} can be regarded as the Laplace transform of (the probability distribution function of)  the random variable $Q_{\xi}:(0,1)\to\Rbb$ under the
probability measure generated by $\overbar{w}$.  
See Appendix \ref{proof:lem:h} for more details about function $h$.  

Similarly, when $\gamma=0$,
the RDU of the terminal wealth $X^{t,\varepsilon,\boldsymbol{\kappa}}(T)$
at time $t$ is 
\begin{align}\label{eq:f:log}
f(\varepsilon;t,\boldsymbol{\kappa})\trieq & J(t,x;\boldsymbol{\pi}^{t,\varepsilon_{0},\boldsymbol{\kappa}})\\ 
= & \ln x+r(T-t)+g(\varepsilon;t,\boldsymbol{\kappa})-\frac{1}{2}H(\varepsilon;t,\boldsymbol{\kappa})-\sqrt{H(\varepsilon;t,\boldsymbol{\kappa})}h^\prime(0),\nonumber
\end{align}
where we use the fact that $h'(0)=-\int_0^1 Q_\xi(p)\dt \overbar w(p)$; see Lemma \ref{lem:h}.

Obviously,  $f(0;t,\bm{\kappa})=J(t,x;\bm{\pi})$ for any $\gamma\in\mathbb{R}$.

Let $\Pi(t)\trieq\int_{t}^{T}\left|\bm{\pi}^{\top}(s)\bm{\sigma}\right|^{2}\dt s$,
$t\in[0,T]$. Obviously, $\Pi(T)=0$. Let 
\[
T_{0}\trieq\inf\left\{ t\in[0,T]:\Pi(t)=0\right\} .
\]
Obviously, $\Pi(t)>0$ for all $t\in[0,T_{0})$ and $\Pi(t)=0$ for
all $t\in[T_{0},T]$. 
Clearly, $\bm{\pi}=\bm{0}$ if and only if $T_0=0$.
Therefore,  $\boldsymbol{\pi}$ is non-zero if and only if $T_{0}>0$.

The following proposition characterizes condition (\ref{eq:mv01:strict}) for $t\in[0,T_0)$.

\begin{prop}
\label{prop:T0>0} Suppose Assumption \ref{assu:DistFun} holds. Given
a strategy $\bm{\pi}\in\Dc$ with $T_{0}>0$, for any $t\in[0,T_{0})$,
condition (\ref{eq:mv01:strict}) holds if and only if 
\begin{numcases}{}
\bm{\mu}-\left[1+\frac{h'\left(-\gamma\sqrt{\Pi(t)}\right)}{h\left(-\gamma\sqrt{\Pi(t)}\right)\sqrt{\Pi(t)}}\right]\bm{\sigma}\bm{\sigma}^{\top}\boldsymbol{\pi}(t)=\boldsymbol{0},\label{eq:k-coe:strict}\\
1+\frac{h'\left(-\gamma\sqrt{\Pi(t)}\right)}{h\left(-\gamma\sqrt{\Pi(t)}\right)\sqrt{\Pi(t)}}>0.\label{eq:k2-coe:strict}
\end{numcases}
\end{prop}

\begin{proof}
We only prove it for the case $\gamma\ne0$, since the proof for the
case $\gamma=0$ is similar. Assume $\gamma\ne0$. Let $t\in[0,T_{0})$.
Then the right-hand derivative of $f$ with respect to $\varepsilon$ is
\begin{align*}
f'_+(\varepsilon;t,\boldsymbol{\kappa})  = & x^{\gamma}\exp\left\{ \gamma\left[r(T-t)+g(\varepsilon;t,\boldsymbol{\kappa})-\frac{1}{2}H(\varepsilon;t,\boldsymbol{\kappa})\right]\right\} h\left(-\gamma\sqrt{H(\varepsilon;t,\boldsymbol{\kappa})}\right)\\
   & \times\left\{ g'_+(\varepsilon;t,\boldsymbol{\kappa})-\frac{1}{2}H'_+(\varepsilon;t,\boldsymbol{\kappa})-\frac{h'\left(-\gamma\sqrt{H(\varepsilon;t,\boldsymbol{\kappa})}\right)H'_+(\varepsilon;t,\boldsymbol{\kappa})}{2h\left(-\gamma\sqrt{H(\varepsilon;t,\boldsymbol{\kappa})}\right)\sqrt{H(\varepsilon;t,\boldsymbol{\kappa})}}\right\} .
\end{align*}
Obviously, the right-hand derivatives
\begin{align*}
g'_+(\varepsilon;t,\boldsymbol{\kappa})  = & \boldsymbol{\kappa}^\top\bm{\mu},\\
H'_+(\varepsilon;t,\boldsymbol{\kappa})  = & \left(2\boldsymbol{\pi}(t+\varepsilon)+\boldsymbol{\kappa}\right)^{\top}\bm{\sigma}\bm{\sigma}^{\top}\boldsymbol{\kappa}.
\end{align*}
Noting that $H(0;t,\boldsymbol{\kappa})=\Pi(t)>0$ for $t\in[0,T_{0})$,
it holds that 
\begin{align*}
  & \lim_{\varepsilon\downarrow0}\frac{f(\varepsilon;t,\boldsymbol{\kappa})-f(0;t,\boldsymbol{\kappa})}{\varepsilon}\\
  = & f'_+(0;t,\boldsymbol{\kappa})\\
  = & x^{\gamma}\exp\left\{ \gamma\left[r(T-t)+g(0;t,\boldsymbol{\kappa})-\frac{1}{2}\Pi(t)\right]\right\} h\left(-\gamma\sqrt{\Pi(t)}\right)\\
   & \times\left\{ \boldsymbol{\kappa}^\top\bm{\mu}-\frac{1}{2}\left(2\boldsymbol{\pi}(t)+\boldsymbol{\kappa}\right)^{\top}\bm{\sigma}\bm{\sigma}^{\top}\boldsymbol{\kappa}
  -\frac{h'\left(-\gamma\sqrt{\Pi(t)}\right)\left(2\boldsymbol{\pi}(t)+\boldsymbol{\kappa}\right)^{\top}\bm{\sigma}\bm{\sigma}^{\top}\boldsymbol{\kappa}}{2h\left(-\gamma\sqrt{\Pi(t)}\right)\sqrt{\Pi(t)}}\right\} \\
  = & \frac{1}{2}x^{\gamma}\exp\left\{ \gamma\left[r(T-t)+g(0;t,\boldsymbol{\kappa})-\frac{1}{2}\Pi(t)\right]\right\} h\left(-\gamma\sqrt{\Pi(t)}\right)G(t,\boldsymbol{\kappa}),
\end{align*}
where 
\begin{align*}
G(t,\boldsymbol{\kappa})  \triangleq & -\left[1+\frac{h'\left(-\gamma\sqrt{\Pi(t)}\right)}{h\left(-\gamma\sqrt{\Pi(t)}\right)\sqrt{\Pi(t)}}\right]\left|\boldsymbol{\kappa}^{\top}\bm{\sigma}\right|^{2}\\
   & +2\left\{ \bm{\mu}^{\top}-\left[1+\frac{h'\left(-\gamma\sqrt{\Pi(t)}\right)}{h\left(-\gamma\sqrt{\Pi(t)}\right)\sqrt{\Pi(t)}}\right]\boldsymbol{\pi}^{\top}(t)\bm{\sigma}\bm{\sigma}^{\top}\right\} \boldsymbol{\kappa}.
\end{align*}
Obviously, (\ref{eq:mv01:strict}) holds 
if and only if $G(t,\boldsymbol{\kappa})<0$ for all  $\boldsymbol{\kappa}\in L^{\infty}(\Fc_{t};\mathbb{R}^{n})\setminus\{\boldsymbol{0}\}$,
which is equivalent to (\ref{eq:k-coe:strict}) and (\ref{eq:k2-coe:strict}). 
\end{proof}

The following proposition investigates condition (\ref{eq:mv01:strict}) for $t\in[T_0,T)$.

\begin{prop}
\label{prop:T0<T}Suppose Assumption \ref{assu:DistFun} holds. Given
a strategy $\boldsymbol{\pi}\in\Dc$ with $T_{0}<T$, the following
assertions hold: 
\begin{description}
\item[(i)] If $h'(0)<0$, then (\ref{eq:mv01:strict}) holds for no $t\in[T_{0},T)$; 
\item[(ii)] If $h'(0)>0$, then (\ref{eq:mv01:strict}) holds for all $t\in[T_{0},T)$;
\item[(iii)] If $h'(0)=0$, then (\ref{eq:mv01:strict}) holds for all $t\in[T_{0},T)$
if and only if 
\begin{equation}
\bm{\mu}=\boldsymbol{0}\text{ and }1-\gamma h''(0)>0.\label{eq:0eqcond}
\end{equation}
\end{description}
\end{prop}

\begin{proof}
We only prove it for the case $\gamma\ne0$, since the proof for the
case $\gamma=0$ is similar. Assume $\gamma\ne0$. Obviously, $\boldsymbol{\pi}(t)=\boldsymbol{0}$
for $t\in[T_{0},T)$.
Then
\begin{align*}
f(\varepsilon;t,\boldsymbol{\kappa})
=  \frac{1}{\gamma}x^{\gamma}e^{\gamma\left[r(T-t)+\zeta(\boldsymbol{\kappa})\varepsilon\right]}h\left(-\gamma\sqrt{\left|\bm{\kappa}^{\top}\bm{\sigma}\right|^{2}\varepsilon}\right),
\end{align*}
where $\zeta(\boldsymbol{\kappa})\trieq\boldsymbol{\kappa}^\top\bm{\mu}-\frac{1}{2}\left|\bm{\kappa}^{\top}\bm{\sigma}\right|^{2}$. 
Obviously, the derivative of $f$ with respect to $\varepsilon$ is
\begin{align*}
f'(\varepsilon;t,\bm{\kappa})= & x^{\gamma}e^{\gamma\left[r(T-t)+\zeta(\boldsymbol{\kappa})\varepsilon\right]}\left[\zeta(\boldsymbol{\kappa})h\left(-\gamma\sqrt{\left|\bm{\kappa}^{\top}\bm{\sigma}\right|^{2}\varepsilon}\right)
 -\sqrt{\left|\bm{\kappa}^{\top}\bm{\sigma}\right|^{2}}\frac{h'\left(-\gamma\sqrt{\left|\bm{\kappa}^{\top}\bm{\sigma}\right|^{2}\varepsilon}\right)}{2\sqrt{\varepsilon}}\right].
\end{align*}

Obviously, $f(\varepsilon;t,\bm{\kappa})-f(0;t,\bm{\kappa})=\bm{0}$ on
$[\bm{\kappa}=\bm{0}]$. Moreover,   $|\bm{\kappa}^\top\bm{\sigma}|^2=\bm{\kappa}^\top\bm{\sigma}\bm{\sigma}^\top\bm{\kappa}>0$ on
$[\bm{\kappa}\ne0]$.

\begin{itemize}
\item[(i)] If $h'(0)<0$, then on $[\bm{\kappa}\neq\bm{0}]$, by the mean value theorem,
\[
\limsup_{\varepsilon\downarrow0}\frac{f(\varepsilon;t,\bm{\kappa})-f(0;t,\bm{\kappa})}{\varepsilon}\ge\liminf_{\varepsilon\downarrow0}f'(\varepsilon;t,\bm{\kappa})=\infty,
\]
which implies that (\ref{eq:mv01:strict}) holds for no $t\in[T_{0},T)$.
\item[(ii)] If $h'(0)>0$, then on $[\bm{\kappa}\neq\bm{0}]$, by the mean value theorem,
\[
\lim_{\varepsilon\downarrow0}\frac{f(\varepsilon;t,\bm{\kappa})-f(0;t,\bm{\kappa})}{\varepsilon}\le\limsup_{\varepsilon\downarrow0}f'(\varepsilon;t,\bm{\kappa})=-\infty,
\]
which implies that (\ref{eq:mv01:strict}) holds for all $t\in[T_{0},T)$.
\item[(iii)] If $h'(0)=0$, then 
\begin{align*}
\lim_{\varepsilon\downarrow0}\frac{f(\varepsilon;t,\bm{\kappa})-f(0;t,\bm{\kappa})}{\varepsilon}  = & \lim_{\varepsilon\downarrow0}f'(\varepsilon;t,\bm{\kappa})\\
  = & x^{\gamma}e^{\gamma r(T-t)}\left[\zeta(\boldsymbol{\kappa})+\frac{1}{2}\gamma\left|\bm{\kappa}^{\top}\bm{\sigma}\right|^{2}h''\left(0\right)\right]\\
  = & x^{\gamma}e^{\gamma r(T-t)}\left[\boldsymbol{\kappa}^\top\bm{\mu}-\frac{1}{2}\left|\bm{\kappa}^{\top}\bm{\sigma}\right|^{2}\left(1-\gamma h''\left(0\right)\right)\right].
\end{align*}
Then (\ref{eq:mv01:strict}) holds for all $t\in[T_{0},T)$ if and only if
(\ref{eq:0eqcond}) holds. \qedhere
\end{itemize}
\end{proof}

Assume that $\bm{\pi}$ is a DSES such that $T_0=T$. 
By Proposition \ref{prop:T0>0}, for all $t\in[0,T)$,
\[
\boldsymbol{\pi}(t)=\left(\bm{\sigma}\bm{\sigma}^{\top}\right)^{-1}\bm{\mu}\left[1+\frac{h'\left(-\gamma\sqrt{\Pi(t)}\right)}{h\left(-\gamma\sqrt{\Pi(t)}\right)\sqrt{\Pi(t)}}\right]^{-1},
\]
which implies 
\[
\left|\boldsymbol{\pi}^{\top}(t)\bm{\sigma}\right|^{2}=\theta^{2}\left[1+\frac{h'\left(-\gamma\sqrt{\Pi(t)}\right)}{h\left(-\gamma\sqrt{\Pi(t)}\right)\sqrt{\Pi(t)}}\right]^{-2},
\]
where $\theta\triangleq\sqrt{\bm{\mu}^{\top}\left(\bm{\sigma}\bm{\sigma}^{\top}\right)^{-1}\bm{\mu}}$.
Recalling the definition of $\Pi$, one knows that $\Pi$
is a positive solution to the following autonomous ODE:
\begin{align}
\begin{cases}
Y'(t)=-\theta^{2}\left[1+\frac{h'\left(-\gamma\sqrt{Y(t)}\right)}{h\left(-\gamma\sqrt{Y(t)}\right)\sqrt{Y(t)}}\right]^{-2}, & t\in[0,T),\\
Y(T)=0.
\end{cases}
\label{eq:Y-ode}
\end{align}
Hereafter, a function $Y$ is called a \emph{positive
solution} to (\ref{eq:Y-ode}) if it satisfies (\ref{eq:Y-ode}) and
is strictly positive on $[0,T)$. For more details about ODE (\ref{eq:Y-ode}), see Appendix \ref{sec:ode}.

\begin{rem}{}
\begin{description}
\item[(i)] Although (\ref{eq:Y-ode}) looks like the ODE (5) of \citet*{hjz2021},
there is a significant difference between them: the former depends
on the agent's utility function while the latter does not.

\item[(ii)] In the case $\gamma=0$, equation (\ref{eq:Y-ode}) reduces to
\begin{align}
\begin{cases}
Y'(t)=-\theta^{2}\left[\frac{\sqrt{Y(t)}}{\sqrt{Y(t)}+h'(0)}\right]^{2}, & t\in[0,T),\\
Y(T)=0.
\end{cases}
\label{eq:Y-ode-log}
\end{align}
\end{description}
\end{rem}

To characterize the DSESes, we need to introduce a function.  
Assuming $h^\prime(0)=0$, let function $\Gc$ be given by
\begin{equation}\label{eq:Gc}
\mathcal{G}(y)\triangleq\int_{0}^{y}\left[1+\frac{h'\left(-\gamma\sqrt{z}\right)}{h\left(-\gamma\sqrt{z}\right)\sqrt{z}}\right]^{2}dz,\quad y\in[0,\infty].
\end{equation}
 Obviously, $\Gc(0)=0$ and $0\le\Gc(y)<\infty$ for all  $y\in[0,\infty)$ since 
$$\lim_{z\downarrow 0}\left(1+\frac{h'\left(-\gamma\sqrt{z}\right)}{h\left(-\gamma\sqrt{z}\right)\sqrt{z}}\right)=\lim_{z\downarrow0}\left(1-\gamma\frac{h'\left(z\right)}{h\left(z\right)z}\right)=1-\gamma h^{\prime\prime}(0).$$

Let
\[
y_{1}\trieq\inf\left\{ y>0\;:\;1+\frac{h'\left(-\gamma\sqrt{y}\right)}{h\left(-\gamma\sqrt{y}\right)\sqrt{y}}\le0\right\} \quad\text{with }\inf\emptyset=\infty
\]
and
\[
\mbb{Y}\triangleq\left\{ y>0\;:\;1+\frac{h'\left(-\gamma\sqrt{y}\right)}{h\left(-\gamma\sqrt{y}\right)\sqrt{y}}>0\right\}.
\]
Obviously, $(0,y_{1})\subseteq\mbb{Y}$. 
If $y_1>0$, then 
$\mathcal{G}:[0,y_1]\to[0,\Gc(y_1)]$  is strictly increasing and 
has an inverse function, denoted by $\mathcal{G}^{-1}$.
In particular, if $\gamma\in(-\infty,0]$, then $y_{1}=\infty$
and $\mathcal{G}(\infty)=\infty$.

The following theorem is the main result of this section, which characterizes the DESEes.

\begin{thm}
\label{thm:dete-eq}Suppose Assumption \ref{assu:DistFun} holds.
The following assertions hold: 
\begin{description}
\item[(i)] If $h'(0)<0$, then there is no DSES; 
\item[(ii)] If $h'(0)>0$, then $\bm{0}$ is the unique DSES; 
\item[(iii)] If $h'(0)=0$, $\bm{\mu}=\bm{0}$, and $1-\gamma h''(0)>0$, then $\bm{0}$ is the unique DSES;
\item[(iv)] If $h'(0)=0$, $\bm{\mu}=\bm{0}$, and $1-\gamma h''(0)\leq0$, then there is no DSES;
\item[(v)] If $h'(0)=0$, 
$\bm{\mu}\ne\bm{0}$, and $\mathcal{G}\left(y_{1}\right)>\theta^{2}T$,
then the following strategy 
\begin{equation}
\boldsymbol{\pi}(t)=\left(\bm{\sigma}\bm{\sigma}^{\top}\right)^{-1}\bm{\mu}\left[1+\frac{h'\left(-\gamma\sqrt{Y(t)}\right)}{h\left(-\gamma\sqrt{Y(t)}\right)\sqrt{Y(t)}}\right]^{-1},\quad t\in[0,T)\label{eq:pi-eq}
\end{equation}
is the unique DSES,
where 
\[Y(t)=\mathcal{G}^{-1}\left(\theta^{2}(T-t)\right),\quad t\in[0,T];\]
\item[(vi)] If $h'(0)=0$,  $\bm{\mu}\ne\bm{0}$, and
$\mathcal{G}\left(y_{1}\right)\leq\theta^{2}T$,
then there is no DSES.
\end{description}
\end{thm}

\begin{proof}
\begin{description}
\item[(i)] Assume $h'(0)<0$. Suppose on the contrary that $\boldsymbol{\pi}$ is a DSES.  Then by Proposition \ref{prop:T0>0}, it satisfies (\ref{eq:k2-coe:strict}).
Moreover,
Proposition \ref{prop:T0<T}(i) implies
$T_{0}=T$. Then the corresponding $\Pi(t)>0$ for all $t\in[0,T)$
and 
\[
\lim_{t\uparrow T}\left[1+\frac{h'\left(-\gamma\sqrt{\Pi(t)}\right)}{h\left(-\gamma\sqrt{\Pi(t)}\right)\sqrt{\Pi(t)}}\right]=-\infty,
\]
which contradicts (\ref{eq:k2-coe:strict}).

\item[(ii)] Assume $h'(0)>0$. Proposition \ref{prop:T0<T}(ii) implies that
$\boldsymbol{0}$ is a DSES.
It is left to show that there is no non-zero DSES.
Suppose on the contrary that there exists a DSES $\boldsymbol{\pi}$ with $T_{0}>0$. 
Then by Proposition \ref{prop:T0>0} and the same way to derive ODE \eqref{eq:Y-ode},
we know that $\Pi$
is a positive solution to 
\begin{equation}
\begin{cases}
Y'(t)=-\theta^{2}\left[1+\frac{h'\left(-\gamma\sqrt{Y(t)}\right)}{h\left(-\gamma\sqrt{Y(t)}\right)\sqrt{Y(t)}}\right]^{-2}, & t\in[0,T_{0}),\\
Y(T_{0})=0.
\end{cases}\label{eq:Y-ode-T_0}
\end{equation}
Similar to Proposition \ref{prop:exist-Y}(i), we have $h'(0)=0$,
which is impossible.

\item[(iii)] Assume $h'(0)=0$, $\bm{\mu}=0$, and $1-\gamma h''(0)>0$. Proposition \ref{prop:T0<T}(iii) implies that
$\boldsymbol{0}$ is a DSES.
It is left to show that there is no non-zero DSES.
Suppose on the contrary that there exists a DSES $\boldsymbol{\pi}\in\Dc$
with $T_{0}>0$. 
Then by Proposition \ref{prop:T0>0},
\begin{equation}\label{eq:pi=0}
\boldsymbol{\pi}(t)=\left(\bm{\sigma}\bm{\sigma}^{\top}\right)^{-1}\bm{\mu}\left[1+\frac{h'\left(-\gamma\sqrt{\Pi(t)}\right)}{h\left(-\gamma\sqrt{\Pi(t)}\right)\sqrt{\Pi(t)}}\right]^{-1}=0\quad\forall\,t\in[0,T_{0}).
\end{equation}
Therefore, $T_0=0$, which is impossible.

\item[(iv)] Assume $h'(0)=0$, $\bm{\mu}=0$, and $1-\gamma h''(0)\leq0$. If there is a DSES $\boldsymbol{\pi}$, then Proposition \ref{prop:T0<T}(iii) implies that $T_0=T$. Consequently, it follows from Proposition \ref{prop:T0>0} that the equality in (\ref{eq:pi=0}) holds for all $t\in[0,T)$. Therefore, $T_0=0$, which is impossible.

\item[(v)] Assume
$h'(0)=0$, $\bm{\mu}\ne\bm{0}$,
and $\mathcal{G}\left(y_{1}\right)>\theta^{2}T$.

First, we verify that (\ref{eq:pi-eq})
gives a DSES. Proposition \ref{prop:exist-Y}(ii) implies that $Y(t)\in\mathbb{Y}$ for all $t\in[0,T)$, that is,
\[
1+\frac{h'\left(-\gamma\sqrt{Y(t)}\right)}{h\left(-\gamma\sqrt{Y(t)}\right)\sqrt{Y(t)}}>0\quad\forall t\in[0,T).
\]
Furthermore, it follows from (\ref{eq:pi-eq}) and Proposition \ref{prop:exist-Y}(ii) that
\[
\Pi'(t)=-\left|\boldsymbol{\pi}^{\top}(t)\bm{\sigma}\right|^{2}=-\theta^{2}\left[1+\frac{h'\left(-\gamma\sqrt{Y(t)}\right)}{h\left(-\gamma\sqrt{Y(t)}\right)\sqrt{Y(t)}}\right]^{-2}=Y'(t)\quad\forall t\in[0,T),
\]
which together with $\Pi(T)=Y(T)=0$ implies that $\Pi=Y$ on $[0,T]$.
Then (\ref{eq:k-coe:strict}) and (\ref{eq:k2-coe:strict}) are satisfied for $t\in[0,T)$. Obviously, $T_0=T$ and
hence, by Proposition \ref{prop:T0>0}, we know that (\ref{eq:mv01:strict})
holds for $t\in[0,T)$. Therefore, (\ref{eq:pi-eq})
gives a DSES. 

Second, we show the uniqueness. 
Let $\boldsymbol{\pi}$ be given by (\ref{eq:pi-eq}),
$\overtilde{\boldsymbol{\pi}}$ be another DSES and $\overtilde{\Pi}(t)\trieq\int_{t}^{T}\left|\overtilde{\boldsymbol{\pi}}^{\top}(s)\bm{\sigma}\right|^{2}\dt s$ for all $t\in[0,T]$.
By Proposition \ref{prop:T0<T}(iii), 
we have $\overtilde{T}_{0}=T$, where $\overtilde{T}_{0}\trieq\inf\left\{ t\in[0,T]:\overtilde{\Pi}(t)=0\right\}$. 
It follows from Proposition \ref{prop:T0>0} that $\overtilde{\Pi}$ is also a positive solution to (\ref{eq:Y-ode}).
Then it follows from Proposition \ref{prop:exist-Y}(ii) that $\Pi=\overtilde{\Pi}$. Consequently, it follows from (\ref{eq:k-coe:strict}) that
$\boldsymbol{\pi}=\overtilde{\boldsymbol{\pi}}$.

\item[(vi)] Assume $h'(0)=0$,
$\bm{\mu}\ne\bm{0}$, and $\mathcal{G}\left(y_{1}\right)\leq\theta^{2}T$. 
Suppose on the contrary that there is a DSES $\boldsymbol{\pi}$.
Proposition \ref{prop:T0<T}(iii) implies the corresponding $T_{0}=T$.
Then Proposition \ref{prop:T0>0} implies that (\ref{eq:k-coe:strict}) and (\ref{eq:k2-coe:strict}) hold on $[0,T)$. 
Since $\mathcal{G}\left(y_{1}\right)\leq\theta^{2}T$, there exits $t_1\in[0,T]$ such that $\theta^2(T-t_1)=\Gc(y_1)$.
Similar to the proof of Proposition \ref{prop:exist-Y}(ii), we have $\Pi(t)=\mathcal{G}^{-1}(\theta^{2}(T-t))$ for $t\in[t_1,T]$. 
Suppose $y_1>0$. Then $t_1<T$. By $y_1=\Pi(t_1)<\infty$, it holds that 
\[
1+\frac{h'\left(-\gamma\sqrt{\Pi(t_1)}\right)}{h\left(-\gamma\sqrt{\Pi(t_1)}\right)\sqrt{\Pi(t_1)}}=1+\frac{h'\left(-\gamma\sqrt{y_1}\right)}{h\left(-\gamma\sqrt{y_1}\right)\sqrt{y_1}}=0.
\]
Therefore, (\ref{eq:k2-coe:strict}) fails at $t_1$, which leads to a contradiction. Thus we must have $y_1=0$.
Then by the definition of $y_1$, for any $\varepsilon>0$, there exits $y_\varepsilon\in(0,\varepsilon)$ such that 
\begin{equation*}
1+\frac{h'\left(-\gamma\sqrt{y_\varepsilon}\right)}{h\left(-\gamma\sqrt{y_\varepsilon}\right)\sqrt{y_\varepsilon}}\le0.
\end{equation*}
By $T_0=T$, we have $\Pi(t)>\Pi(T)=0$ for all $t\in[0,T)$.  
Then for any $\varepsilon\in(0,\Pi(0))$,  
there exits $t_2\in(0,T)$ such that $\Pi(t_2)=y_\varepsilon$. Therefore, $(\ref{eq:k2-coe:strict})$ does not hold at $t_2$, which is a contradiction.\qedhere
\end{description} 
\end{proof}

Theorem \ref{thm:dete-eq} shows that, in any case, there is up to one DSES if the probability weighting function is time-invariant and satisfies Assumption \ref{assu:DistFun}. 

\begin{rem}
Consider the case $\gamma=0$ and $h'(0)=0$. By assertions (iii) and (v)
of Theorem \ref{thm:dete-eq}, the unique DSES is 
\begin{equation*}
\boldsymbol{\pi}(t)=\left(\boldsymbol{\sigma}\boldsymbol{\sigma}^{\top}\right)^{-1}\bm{\mu},\quad t\in[0,T),
\end{equation*}
which is independent of the distortion function and coincides with
the solution of \citet*{m71}.
\end{rem}

\begin{example}\label{exm:w:Phi}
Let $\Phi$ be the standard normal distribution function.
Consider a class of probability weighting functions $w$ defined by
\begin{equation*}
w(p)=\Phi\left(\frac{1}{\lambda}\left(\Phi^{-1}(p)+\nu\right)\right),\quad p\in[0,1],
\end{equation*}
where $\lambda>0$ and $\nu\in\mathbb{R}$ are two parameters.
Obviously, if $\lambda=1$ and $\nu=0$, then $w$
is the identity function. If $\lambda=1$ and $\nu\neq0$,
then $w$ is the distortion function discussed by \citet*{w2000}; see also \citet*{hsz2021} for an application. 
If $\lambda=\sqrt{2}$ and $\nu=0$,
then $w$ is the distortion function considered by
\citet*[Section 5]{hjz2021}.
For this class of probability weighting functions $w$, we have
$$w'(p)=\frac{\Phi'\left(\frac{1}{\lambda}\left(\Phi^{-1}(p)+\nu\right)\right)}{\lambda\Phi'(\Phi^{-1}(p))},\quad p\in(0,1).$$
Therefore,
\begin{equation}\label{eq:wp:Phi}
w'(\Phi(z))=\frac{\Phi'\left(\frac{1}{\lambda}\left(z+\nu\right)\right)}{\lambda\Phi'(z)},\quad z\in\mathbb{R}.
\end{equation}
That is 
\begin{align*}
w'(\Phi(z))=&{1\over\lambda}\exp\left\{-{(z+\nu)^2\over 2\lambda^2}+{z^2\over2}\right\}\\
=&{1\over2}\exp\left\{{1\over2\lambda^2}\left[(\lambda^2-1)z^2-2\nu z-\nu^2\right]\right\},\quad z\in\mathbb{R}.
\end{align*}
\begin{itemize}
\item Assume $\lambda>1$. In this case, $w'(\Phi(\cdot))$ is strictly decreasing on $(-\infty,z_0)$ and strictly increasing on $(z_0,\infty)$, where $z_0=\nu/(\lambda^2-1)$. Then, $w'$ is strictly decreasing on $(0,\Phi(z_0))$ and strictly 
increasing on $(\Phi(z_0),1)$. That is, $w$ is strictly concave on $(0,\Phi(z_0))$ and strictly 
convex on $(\Phi(z_0),1)$. Therefore, $w$ is inverse-S shaped. 
\item Similarly, if $\lambda\in(0,1)$, then $w$ is S-shaped; 
\item If $\lambda=1$ and $\nu>0$, then $w$ is concave;
\item If $\lambda=1$ and $\nu<0$, then $w$ is convex.
\end{itemize}
By \eqref{eq:wp:Phi}, in a similar way of \citet*[Section 5]{hjz2021}, we can verify that this class of probability weighting functions satisfy Assumption \ref{assu:DistFun}.
The function $h$ can be calculated as follows (by \eqref{eq:wp:Phi} and \eqref{eq:fn-h}):
\begin{align*}
h(x) = & \Ebf\left[e^{x\xi}w'\left(\Phi(\xi)\right)\right]\\
=&{1\over\lambda}\int_{-\infty}^\infty e^{xz}\Phi'\left({1\over\lambda}(z+\nu)\right)\dt z\\
=&\int_{-\infty}^\infty e^{x(\lambda y-\nu)}\Phi'(y)\dt y\\
=&\Ebf[e^{x(\lambda\xi-\nu)}]\\
= & e^{-\nu x+\frac{\lambda^{2}}{2}x^{2}},\quad x\in\mathbb{R}.
\end{align*}
Therefore, $h$ is the Laplace transform of the Normal distribution $N(\nu,\lambda^{2})$.
Obviously,
\begin{align*}
h'(x)  = & \left(\lambda^{2}x-\nu\right)h(x),\\
h^{\prime\prime}(x)  = & \left[\lambda^{2}+\left(\lambda^{2}x-\nu\right)^{2}\right]h(x).
\end{align*}
In particular, $-h'(0)=\nu$, $h^{\prime\prime}(0)=\lambda^2+\nu^2$ and
$1-\gamma h^{\prime\prime}(0)=1-\gamma(\lambda^2+\nu^2)$. Moreover,  if $\nu=0$, then
$$1+\frac{h'\left(-\gamma\sqrt{z}\right)}{h\left(-\gamma\sqrt{z}\right)\sqrt{z}}=1-\gamma\lambda^2,\quad z\in(0,\infty).$$ 
Now applying Theorem \ref{thm:dete-eq} and Propositions \ref{prop:T0>0}--\ref{prop:T0<T} yields the following assertions.

\begin{itemize}
\item If $\nu>0$, then there exits no DSES.
\item If $\nu<0$, then $\bm{0}$ is the unique DSES.
\item If $\nu=0$, $1-\gamma\lambda^2>0$ and $\bm{\mu}=\bm{0}$, then $\bm{0}$ is the unique DSES.

\item If $\nu=0$, $1-\gamma\lambda^2>0$, $\bm{\mu}\ne0$, then  $y_1=\infty$ and 
$$\Gc(y)=\int_0^y(1-\gamma\lambda^2)^2\dt z=(1-\gamma\lambda^2)^2y,\quad y\in[0,\infty).$$
Therefore, $\Gc(y_1)=\infty>\theta^2T$ and the unique DSES (\ref{eq:pi-eq})
is a constant vector given by
\[
\boldsymbol{\pi}(t)=\left(\bm{\sigma}\bm{\sigma}^{\top}\right)^{-1}\bm{\mu}\left(1-\gamma\lambda^{2}\right)^{-1},\quad t\in[0,T).
\]
On the other hand, ODE (\ref{eq:Y-ode}) becomes
\[
\begin{cases}
Y'(t)=-\theta^{2}\left(1-\gamma\lambda^{2}\right)^{-2}, & t\in[0,T),\\
Y(T)=0,
\end{cases}
\]
which admits a unique solution $Y(t)=\theta^{2}\left(1-\gamma\lambda^{2}\right)^{-2}(T-t)$.
\item If $\nu=0$ and $1-\gamma\lambda^2=0$, 
then $y_1=0$ and there is no  DSES.
\end{itemize}
\end{example}

\section{Time-Dependent Probability Weighting Functions}\label{sec:td}

In this section, we consider the case where the probability weighting
function depends on time. The time-$t$ RDU $\ensuremath{\Ebf_{t}^{w}[U(X)]}$
is now given by 
\begin{equation*}
\ensuremath{\Ebf_{t}^{w}[U(X)]}=\int_{0}^{1}U\left(Q_{X}^{t}(p)\right)\overbar{w}(t,\dt p),
\end{equation*}
where $\overbar{w}(t,p)\trieq1-w(t,p)$ for all $(t,p)\in[0,T)\times[0,1]$
with each $w(t,\cdot)$ being a probability weighting
function and the integration is calculated under the Lebesgue-Stieltjes measure generated by $\overbar  w(t,\cdot)$. 
We impose the following assumption.

\begin{assumption}
\label{assu:DistFun-t} $w:[0,T)\times[0,1]\to[0,1]$ is measurable
and, for each $t\in[0,T)$, $w(t,\cdot)$ satisfies Assumption \ref{assu:DistFun}. 
\end{assumption}

In the following, the first, second and third order
partial derivatives w.r.t. the second argument of a function
$f:[0,T]\times\mathbb{R}\to\mathbb{R}$ are denoted by $f_{x}$, $f_{xx}$
and $f_{xxx}$, respectively. 

The time-dependent version of  (\ref{eq:h}) is given by
\begin{equation*}
h(t,x)\trieq\int_{0}^{1}e^{-xQ_{\xi}(p)}\overbar{w}(t, \dt p)
\quad(t,x)\in[0,T)\times\mathbb{R}.
\end{equation*}
Obviously, under Assumption \ref{assu:DistFun-t}, for
each $t\in[0,T)$, the conclusions of Lemma \ref{lem:h} hold for
$h(t,\cdot)$. In particular, for every $t\in[0,T)$, $h(t,0)=1$, $h(t,x)$ is strictly convex in $x$ and $h_x(t,x)$ is strictly increasing in $x$.
If $h_x(t,0)\ge0$, then $h_x(t,x)>h_x(t,0)\ge0$ for all $x\in(0,\infty)$ and hence $h(t,x)> h(t,0)=1$ for all $x\in(0,\infty)$.

In this section, we still focus on the CRRA utility \eqref{eq:Upower} .

A strategy $\boldsymbol{\pi}$ is called \emph{admissible}
if $\int_{0}^{1}\left|U\left(Q_{X^{\boldsymbol{\pi}}(T)}^{t}(p)\right)\right|\overbar{w}(t,\dt p)<\infty$
for every $t\in[0,T)$. We still abuse $\Ac$ to denote the set of all admissible
strategies. The notation $J(t,x;\bm{\pi})$ and the notion of strict equilibrium strategy can be defined in a similar way as in Section \ref{sec:rdu}. We still have $\Dc\subset\Ac$ and  $\boldsymbol{\pi}^{t,\varepsilon,\boldsymbol{\kappa}}\in\Ac$
for all $t\in[0,T)$, $\varepsilon\in(0,T-t)$ and $\boldsymbol{\kappa}\in L^{\infty}(\Fc_{t};\Rbb^{n})$.
Therefore, $\boldsymbol{\pi}\in\Dc$ is a DSES if and only
if, for every $t\in[0,T)$, 
condition \eqref{eq:mv01:strict} is satisfied.

Let $\bm{\pi}\in\Dc$ be fixed and let $\Pi$ and $T_0$ be defined by the same way as in Section \ref{sec:dtmn}.
The following two propositions extend
Propositions  \ref{prop:T0>0}--\ref{prop:T0<T}. The proofs are
similar.

\begin{prop}
\label{prop:T0>0-t} Suppose Assumption \ref{assu:DistFun-t} holds.
Given a strategy $\boldsymbol{\pi}\in\Dc$ with $T_{0}>0$,
for any $t\in[0,T_{0})$, condition (\ref{eq:mv01:strict}) holds if and only if
\begin{numcases}{}
\bm{\mu}-\left[1+\frac{h_{x}\left(t,-\gamma\sqrt{\Pi(t)}\right)}{h\left(t,-\gamma\sqrt{\Pi(t)}\right)\sqrt{\Pi(t)}}\right]\bm{\sigma}\bm{\sigma}^{\top}\boldsymbol{\pi}(t)=\boldsymbol{0},\label{eq:k-coe-t}\\
1+\frac{h_{x}\left(t,-\gamma\sqrt{\Pi(t)}\right)}{h\left(t,-\gamma\sqrt{\Pi(t)}\right)\sqrt{\Pi(t)}}>0.\label{eq:k2-coe-t}
\end{numcases}
\end{prop}

\begin{prop}
\label{prop:T0<T-t}Suppose Assumption \ref{assu:DistFun-t} holds.
Given a strategy $\boldsymbol{\pi}\in\Dc$ with $T_{0}<T$,
for any $t\in[T_{0},T)$, the following assertions hold: 
\begin{description}
\item[(i)] If $h_{x}(t,0)<0$, then (\ref{eq:mv01:strict}) does not hold; 
\item[(ii)] If $h_{x}(t,0)>0$, then (\ref{eq:mv01:strict}) holds; 
\item[(iii)] If $h_{x}(t,0)=0$, then (\ref{eq:mv01:strict}) holds if and only if 
\begin{equation*}
\boldsymbol{\mu}=\boldsymbol{0}\text{ and }1-\gamma h_{xx}(t,0)>0.  
\end{equation*}
\end{description}
\end{prop}

In the case $\gamma\le0$, for every $t\in[0,T)$, condition \eqref{eq:k2-coe-t} automatically holds if $h_x(t,0)\ge0$.
In the case $\gamma>0$,  however, it is quite complicate to guarantee the validity of condition \eqref{eq:k2-coe-t}. 
Hereafter, we focus on the case $\gamma\le0$.  

If $\bm{\pi}$ is a DSES such that $T_0=T$, 
then, based on Proposition \ref{prop:T0>0-t}, by the same way to derive ODE \eqref{eq:Y-ode}, we know that $\Pi$ solves the following ODE:
\begin{equation}
\begin{cases}
Y'_+(t)=-\theta^{2}m^{2}\left(t,\sqrt{Y(t)}\right), & t\in[0,T),\\
Y(T)=0,
\end{cases}\label{eq:Y-ode-t}
\end{equation}
where $Y'_+$ denotes the right-hand derivative of $Y$ and 
\[
m(t,x)\triangleq\frac{h\left(t,-\gamma x\right)x}{h\left(t,-\gamma x\right)x+h_{x}\left(t,-\gamma x\right)},\quad (t,x)\in[0,T)\times(0,\infty).
\]
We extend the definition of $m$ by letting
\begin{equation*}
m(t,0)\trieq\lim_{x\downarrow0}m(t,x)=
\begin{cases}
\frac{1}{1-\gamma h_{xx}(t,0)},\quad &\text{if }h_x(t,0)=0,\\
0,\quad&\text{if }h_x(t,0)>0,
\end{cases}
\qquad t\in[0,T).
\end{equation*}

Let $AC([0,T])$ denotes the class of absolutely continuous functions on $[0,T]$, $C^1_+([0,T))$ the class of right-hand derivable functions with the right-hand derivative being right-continuous on $[0,T)$, $C^1_-((0,T])$ the class of left-hand derivable functions with the left-hand derivative being left-continuous on $(0,T]$. 

\begin{defn} 
A function $Y:[0,T]\to[0,\infty)$ is called a solution to ODE \eqref{eq:Y-ode-t} if $Y\in AC([0,T])\cap C^1_+([0,T))$ and $Y$ satisfies \eqref{eq:Y-ode-t}.
A solution $Y$ to ODE \eqref{eq:Y-ode-t} is called positive if $Y>0$ on $[0,T)$.  A positive solution $\overbar{Y}$ to ODE (\ref{eq:Y-ode-t}) is called maximal if $\overbar{Y}\ge Y$ for every positive solution $Y$ to ODE (\ref{eq:Y-ode-t}).
\end{defn}

If $h$ and $h_x$ are continuous on $[0,T)\times(0,\infty)$, then $m$ is continuous on $[0,T)\times(0,\infty)$ and hence each positive solution to ODE \eqref{eq:Y-ode-t} is continuously differentiable on $[0,T)$. 
Moreover, if $h$ and $h_x$ are continuous on $[0,T)\times[0,\infty)$ and $h_x(t,0)>0$ for all $t\in[0,T)$, then $m$ is continuous on $[0,T)\times[0,\infty)$ and hence each solution to ODE \eqref{eq:Y-ode-t} is continuously differentiable on $[0,T)$. 
In the case of continuously differentiable solution, we will occasionally write the ODE in the following form: 
\begin{equation*}
\begin{cases}
Y'(t)=-\theta^{2}m^{2}\left(t,\sqrt{Y(t)}\right), & t\in[0,T),\\
Y(T)=0.
\end{cases}
\end{equation*}

\begin{rem}
In the case $\gamma=0$, we have
\[
m(t,x)=\frac{x}{x+h_x(t,0)},\quad (t,x)\in[0,T)\times(0,\infty)
\]
and hence
ODE \eqref{eq:Y-ode-t} reads
\begin{align}
\begin{cases}
Y'_+(t)=-\frac{\theta^{2}}{\left(\sqrt{Y(t)}+h_x(t,0)\right)^{2}}Y(t), & t\in[0,T),\\
Y(T)=0.
\end{cases}
\label{eq:Y-ode-log-t}
\end{align}
If $h_x(\cdot,0)$ is continuous on $[0,T)$, then each positive solution to ODE \eqref{eq:Y-ode-log-t} is continuously differentiable.
\end{rem}

The cases $\gamma<0$ and $\gamma=0$ are respectively discussed in the following two subsections. 
Correspondingly, the ODE \eqref{eq:Y-ode-t} with $\gamma<0$ is studied in Appendix \ref{sec:ode-t}, whereas ODE \eqref{eq:Y-ode-log-t} is studied in Appendix \ref{app:ode-log}.

\subsection{Power Utility ($\gamma<0$)}

We focus on the case $\gamma<0$ in this subsection. 

The following theorem investigates the zero equilibrium. 

\begin{thm}\label{thm:dete-eq-t-0} 
Let $\gamma<0$.
Under Assumption \ref{assu:DistFun-t}, we
have the following assertions: 
\begin{description}
\item[(i)] If $h_{x}(t,0)<0$ for some $t\in[0,T)$, then $\bm{0}$ is
not a DSES. If 
\begin{equation}\label{eq:linf_hx}
\liminf_{t\uparrow T}h_x(t,0)<0,\ \liminf_{t\uparrow T,\, x\downarrow 0}h_{x}(t,x)<0,\text{ and }\limsup_{t\uparrow T,\,x\downarrow0}h(t,x)<\infty,
\end{equation}
then there is no DSES; 
\item[(ii)] If $h_{x}(t,0)>0$ for all $t\in[0,T)$, then $\bm{0}$ is
DSES. If $h$ and $h_x$ are continuous on $[0,T)\times[0,\infty)$ and $\inf_{t\in[0,T)}h_{x}(t,0)>0$, then $\bm{0}$ is the unique DSES. 
\end{description}
\end{thm}

\begin{proof}
\begin{description}
\item[(i)] The first part of this assertion is a consequence of Proposition \ref{prop:T0<T-t}(i). We are going to prove the second part. 
Suppose on the contrary that $\boldsymbol{\pi}$ is a DSES.
The first condition in \eqref{eq:linf_hx} and Proposition \ref{prop:T0<T-t}(i) imply that $T_{0}=T$. Then by Proposition \ref{prop:T0>0-t}(ii), condition (\ref{eq:k2-coe-t}) holds for all $t\in[0,T)$. On the other hand, the last two conditions in \eqref{eq:linf_hx} imply 
\[
\liminf_{t\uparrow T}\left[1+\frac{h_x\left(t,-\gamma\sqrt{\Pi(t)}\right)}{h\left(t,-\gamma\sqrt{\Pi(t)}\right)\sqrt{\Pi(t)}}\right]=-\infty.
\]
Then condition (\ref{eq:k2-coe-t}) does not hold for $t$ near to $T$. So we have a contradiction.

\item[(ii)] The first part of this assertion is a consequence of Proposition \ref{prop:T0<T-t}(ii). Now we are going to prove the second part. Suppose on the contrary that there is a DSES $\boldsymbol{\pi}\in\Dc$
with $T_{0}>0$. Then similar to (\ref{eq:Y-ode-T_0}), 
the following ODE
\begin{equation*}
\begin{cases}
Y'_+(t)=-\theta^{2}m^{2}\left(t,\sqrt{Y(t)}\right), & t\in[0,T_0),\\
Y(T_0)=0
\end{cases}
\end{equation*}
admits a solution $Y$ with $Y>0$ on $[0,T_0)$, which is impossible by Proposition \ref{prop:exist-Y-t-0}.
Thus $\bm{0}$ is the unique DSES.\qedhere
\end{description}
\end{proof}

Now we are going to investigate the non-zero DSESes. To this end, we introduce the following assumption.

\begin{assumption}\label{assu:h-t} 
\begin{description}
\item[(i)] Both of $h$ and $h_{x}$ are continuous on $[0,T)\times[0,\infty)$;
\item[(ii)] $h_{x}(t,0)\geq0$  for all $t\in[0,T)$;
\item[(iii)] $\limsup_{t\downarrow0}\frac{h_{x}(T-t,0)}{\theta\sqrt{t}}<1$;
\item[(iv)] $\limsup_{t\downarrow 0, x\downarrow0}h_{xx}(T-t,x)<\infty$.
\end{description}
\end{assumption}

\begin{rem}
In addition to conditions (ii)--(iii) 
in Assumption \ref{assu:h-t},  the following conditions are imposed in \citet*[Assumption 4.2]{hjz2021}:
\begin{description}
\item[(a)] $h_{xxx}(t,0)\geq0$ for all $t\in[0,T]$; 
\item[(b)] $\sup_{t\in[0,T]}h(t,1)<\infty$ and $\limsup_{t\uparrow T}h_{xx}(t,1)<\infty$; 
\item[(c)] $\inf_{t\in[0,T]}h_{xx}(t,0)>0$. 
\end{description}
Condition (a) implies that $h_{xx}(t,x)$ is increasing in $x$ and hence conditions (a)--(b) imply 
Assumption \ref{assu:h-t}(iv). Therefore,  our Assumption \ref{assu:h-t} is weaker than \citet*[Assumption 4.2]{hjz2021}  (except for the continuity of $h$ and $h_x$).
\end{rem}

\begin{thm}\label{thm:dete-eq-t-pstv} 
Let $\gamma<0$ and $\bm{\mu}\ne0$. Under Assumptions \ref{assu:DistFun-t} and \ref{assu:h-t},  we
have the following assertions: 
\begin{description}
\item[(i)] ODE (\ref{eq:Y-ode-t}) has a positive solution $Y$ and
\begin{equation}
\boldsymbol{\pi}(t)=\left(\bm{\sigma}\bm{\sigma}^{\top}\right)^{-1}\bm{\mu}\left[1+\frac{h_{x}\left(t,-\gamma\sqrt{Y(t)}\right)}{h\left(t,-\gamma\sqrt{Y(t)}\right)\sqrt{Y(t)}}\right]^{-1},\quad t\in[0,T)\label{eq:pi-eq-t}
\end{equation}
gives a non-zero DSES;

\item[(ii)] If $h_{x}(t,0)>0$ for all $t\in[0,T)$, then $\boldsymbol{\pi}$ is a non-zero DSES if and only if $\boldsymbol{\pi}$ is given by  \eqref{eq:pi-eq-t} with $Y$ being a positive solution to ODE \eqref{eq:Y-ode-t}.
\end{description}
\end{thm}

\begin{proof}
\begin{description}
\item[(i)] 
Proposition \ref{prop:exist-Y-t} implies that ODE (\ref{eq:Y-ode-t}) has a positive solution. The proof of the rest part is similar to the first part of the proof of Theorem \ref{thm:dete-eq}(v).

\item[(ii)] By assertion (i), ODE \eqref{eq:Y-ode-t} admits positive solutions and if $Y$ is a positive solution to ODE \eqref{eq:Y-ode-t} then \eqref{eq:pi-eq-t} gives a non-zero DSES. Conversely, Let $\bm{\pi}$ be a non-zero DSES. Then $T_0\in(0,T]$.
By the same way to derive ODE \eqref{eq:Y-ode}, we know that
$\Pi$ is a positive solution to the following ODE
\begin{equation*}
\begin{cases}
\Pi'_+(t)=-\theta^{2}m^{2}\left(t,\sqrt{\Pi(t)}\right), & t\in[0,T_0),\\
\Pi(T_0)=0.
\end{cases}
\end{equation*}
Suppose $T_0\in(0,T)$. Then Proposition \ref{prop:exist-Y-t-0} implies that $\Pi(t)=0$ for $t\in[0,T_0]$, which contradicts the definition of $T_0$. Therefore, $T_0=T$ and hence $\Pi$ is a positive solution to ODE \eqref{eq:Y-ode-t}.
Proposition \ref{prop:T0>0-t} implies that
 $\bm{\pi}$ is given by \eqref{eq:pi-eq-t} with $Y=\Pi$.\qedhere
 \end{description}
\end{proof}

 Let $\gamma<0$ and $\bm{\mu}\ne0$. Under Assumptions \ref{assu:DistFun-t} and \ref{assu:h-t}, if $h_x(t,0)>0$ for all $t\in[0,T)$, then by Theorem \ref{thm:dete-eq-t-pstv}(ii), looking for all non-zero DSESes  is equivalent to looking for all positive solutions to ODE \eqref{eq:Y-ode-t}. 
Moreover, by $h_x(t,0)>0$ for all $t\in[0,T)$, we have $m(t,0)=0$ for all $t\in[0,T)$. In this case, $0$ is also a solution to ODE \eqref{eq:Y-ode-t}. This makes it difficult to find a positive solution numerically.  In fact,  when we try the standard numerical method (such as the built-in function NDSolve in Mathematica) to solve ODE \eqref{eq:Y-ode-t} backwardly with the terminal condition $Y(T)=0$, the program is interrupted immediately because zero is input into the denominator and causes the computation to encounter infinity. If there are various positive solutions, it is difficult (if not impossible) to find all positive solutions by solve ODE \eqref{eq:Y-ode-t} backwardly. We can approximate the maximal positive solution by the solution to the backward ODE with the terminal condition $Y(T)=\varepsilon>0$ with $\varepsilon$ sufficiently small. However, we can not find out the other positive solutions in this way. This motives us to look for a ``forward” method to get all positive solutions, which is provided in Appendix \ref{app:FB}. Here we present it briefly.  
The following assumption is needed.
\begin{assumption}\label{ass:zeta}
\begin{description}
\item[(i)] $h_x(t,0)>0$ for all $t\in[0,T)$;
\item[(ii)] $\sup\{\sqrt{x}h_{xx}(t,\sqrt{x}): t\in[0,\delta],x\in[0,\zeta]\}<\infty$ for all $\delta\in(0,T)$ and $\zeta\in(0,\infty)$;
\item[(iii)] For all $\zeta\in(0,\infty)$, $\sup_{t\in[0,T)}h(t,\zeta)<\infty$,
$\inf_{t\in[0,T)}h_x(t,\zeta)>0$, $\sup_{t\in[0,T)}h_x(t,\zeta)<\infty$, and $\sup_{t\in[0,T),x\in(0,\zeta]}h_{xx}(t,x)<\infty$.
\end{description}
\end{assumption}
\noindent Under the previous assumption, by Proposition \ref{prop:exist-Y-t-m}, $Y$ is a positive solution to the backward ODE \eqref{eq:Y-ode-t} if and only if $Y$ solves the following forward ODE 
\begin{equation}
\begin{cases}
Y'(t)=-\theta^{2}m^{2}\left(t,\sqrt{Y(t)}\right), & t\in[0,T),\\
Y(0)=\eta
\end{cases}\label{eq:Y-ode-t-forward}
\end{equation}
for some $\eta\in(0,\eta^*]$, where 
\begin{equation}\label{eq:eta*}
\eta^*\triangleq\sup\{Y(0):Y\text{ is a positive solution to (\ref{eq:Y-ode-t})}\}.
\end{equation}
The forward ODE \eqref{eq:Y-ode-t-forward} admits a unique solution, which can be determined by the standard numerical method. 
Moreover, $\eta^*$ can be determined by the approximation method provided by Remark \ref{rem:approximate eta}, 
that is, solve the following backward ODE
\begin{equation}\label{eq:bode:Y;epsilon}
\begin{cases}
Y'(t)=-\theta^{2}m^{2}\left(t,\sqrt{Y(t)}\right), & t\in[0,T),\\
Y(T)=\varepsilon
\end{cases}
\end{equation}
for sufficiently small $\varepsilon>0$ and the resultant initial value $Y(0)$ is an approximations of $\eta^*$.

The next theorem establishes the uniqueness of the DSES.

\begin{thm}\label{thm:dete-eq-t-unq} 
Let $\gamma<0$ and $\bm{\mu}\ne0$. 
Under Assumptions \ref{assu:DistFun-t},  \ref{assu:h-t}(i,ii), \ref{ass:zeta}(ii),
if 
\begin{equation}\label{eq:cond:h:delta}
\begin{split}
&\text{there exists some }\delta\in(0,T)\text{ such that }h_{x}(t,0)=0 \text{ for all }t\in[T-\delta,T),\\ 
&\sup_{t\in[\delta,T)}h_{xx}(t,0)<\infty,\text{ and }\sup_{(t,x)\in[\delta,T)\times[0,\theta^2\delta]}\left|h_{xxx}(t,x)\right|<\infty,
\end{split}
\end{equation}
then there is a unique DSES, which is given by (\ref{eq:pi-eq-t}) with $Y$
being the unique positive solution to ODE (\ref{eq:Y-ode-t}).
\end{thm}

\begin{proof}
Let $\bm{\pi}$ be given by (\ref{eq:pi-eq-t}) with $Y$
being the unique positive solution to (\ref{eq:Y-ode-t}).
Similar to Theorem \ref{thm:dete-eq-t-pstv}(i), one can verify that $\boldsymbol{\pi}$
is a DSES. Now we show the uniqueness. Let $\overtilde{\boldsymbol{\pi}}$
be a DSES, $\overtilde{\Pi}(t)\trieq\int_{t}^{T}\left|\overtilde{\boldsymbol{\pi}}^{\top}(s)\bm{\sigma}\right|^{2}\dt s$
and $\overtilde{T}_{0}\trieq\inf\left\{ t\in[0,T]:\overtilde{\Pi}(t)=0\right\} $.
Recalling $\boldsymbol{\mu}\neq0$, it follows from Proposition \ref{prop:T0<T-t}(iii)
and condition \eqref{eq:cond:h:delta} that $\overtilde{T}_{0}=T$. 
By the same way to derive ODE \eqref{eq:Y-ode}, we know that $\overtilde{\Pi}$ is a
positive solution to ODE (\ref{eq:Y-ode-t}). By Proposition
\ref{prop:exist-Y-t-u}, ODE (\ref{eq:Y-ode-t}) has a unique solution. Therefore, $\overtilde{\Pi}=Y=\Pi$. 
Then
by Proposition \ref{prop:T0>0-t}, 
we know that (\ref{eq:k-coe-t}) is satisfied by both $\boldsymbol{\pi}$
and $\overtilde{\boldsymbol{\pi}}$. Then we have $\boldsymbol{\pi}=\overtilde{\boldsymbol{\pi}}$.
\end{proof}

\subsection{Logarithmic Utility ($\gamma=0$)}
In this subsection, we present the results for logarithmic utility. The proofs are omitted since they are similar to their counterparts under power utility, based on the results on ODE (\ref{eq:Y-ode-log-t}) in Appendix \ref{app:ode-log}.

We first state the results on zero equilibrium. 
\begin{thm}\label{thm:dete-eq-t-0-log} 
Let $\gamma=0$.
Under Assumption \ref{assu:DistFun-t}, we
have the following assertions: 
\begin{description}
\item[(i)] If $\liminf_{t\uparrow T}h_{x}(t,0)<0$, then there is no DSES; 
\item[(ii)] If $h_{x}(t,0)>0$ for all $t\in[0,T)$, then $\bm{0}$ is
an SES. If  $h_x(\cdot,0)$ is continuous on $[0,T)$ and $\inf_{t\in[0,T)}h_{x}\left(t,0\right)>0$, 
then $\bm{0}$ is the unique DSES. 
\end{description}
\end{thm}

Now we present the results on the non-zero DSESes.

\begin{thm}\label{thm:dete-eq-t-log} 
Let $\gamma=0$ and $\bm{\mu}\ne0$.  In addition to Assumption \ref{assu:DistFun-t}, let $h_{x}(\cdot,0)$ be continuous on $[0,T)$. 
\begin{description}
\item[(i)] If $h_{x}(\cdot,0)\geq0$ on $[0,T)$ and $\limsup_{t\uparrow T}\frac{h_{x}(t,0)}{\theta\sqrt{T-t}}<1$, then ODE (\ref{eq:Y-ode-log-t}) has a positive solution $Y$ and
\begin{equation}
\boldsymbol{\pi}(t)=\left(\bm{\sigma}\bm{\sigma}^{\top}\right)^{-1}\bm{\mu}\left[1+\frac{h_{x}\left(t,0\right)}{\sqrt{Y(t)}}\right]^{-1},\quad t\in[0,T)\label{eq:pi-eq-t-log}
\end{equation}
gives a non-zero DSES;

\item[(ii)] If $h_{x}(\cdot,0)>0$ on $[0,T)$ and $\limsup_{t\uparrow T}\frac{h_{x}(t,0)}{\theta\sqrt{T-t}}<1$, then $\boldsymbol{\pi}$ is a non-zero DSES if and only if $\boldsymbol{\pi}$ is given by  \eqref{eq:pi-eq-t-log} with $Y$ being a positive solution to ODE \eqref{eq:Y-ode-log-t}.
Moreover, $Y$ is a positive solution to the backward ODE \eqref{eq:Y-ode-log-t} if and only if $Y$ solves the following forward ODE 
\begin{equation}
\begin{cases}
Y'(t)=-\frac{\theta^{2}}{\left(\sqrt{Y(t)}+h_x(t,0)\right)^{2}}Y(t), & t\in[0,T),\\
Y(0)=\eta
\end{cases}\label{eq:Y-ode-log-t-forward}
\end{equation}
for some $\eta\in(0,\eta^*]$, where 
$\eta^*\triangleq\sup\{Y(0):Y\text{ is a positive solution to (\ref{eq:Y-ode-log-t})}\}$.

\item[(iii)] If $h_{x}(\cdot,0)\geq0$ on $[0,T)$ and there exists $\delta\in(0,T)$
such that $h_{x}(t,0)=0$ for all $t\in[T-\delta,T)$,  then there is a unique DSES, which is given by (\ref{eq:pi-eq-t-log}) with $Y$
being the unique positive solution to (\ref{eq:Y-ode-log-t}).
\end{description}
\end{thm}

\subsection{Optimal Equilibrium}

We begin with the logarithmic utility case. 

Let $\gamma=0$, $\bm{\mu}\ne0$ and Assumption \ref{assu:DistFun-t} hold.
Assume that $h_{x}(\cdot,0)$ is continuous on $[0,T)$, $h_{x}(\cdot,0)>0$ on $[0,T)$ and $\limsup_{t\uparrow T}\frac{h_{x}(t,0)}{\theta\sqrt{T-t}}<1$. By Theorem \ref{thm:dete-eq-t-log}(ii), 
$\boldsymbol{\pi}$ is a non-zero DSES if and only if $\boldsymbol{\pi}$ is given by  \eqref{eq:pi-eq-t-log} with $Y$ being the unique solution to the forward ODE \eqref{eq:Y-ode-log-t-forward} for some $\eta\in(0,\eta^*]$.
Moreover, $\bm{0}$ is also a DSES, which can be generated by the solution of the forward ODE \eqref{eq:Y-ode-log-t-forward} with $\eta=0$. Therefore, each $\eta\in[0,\eta^*]$ yields a DSES. 
A question arises: which one should the agent choose from these equilibrium strategies? 

To answer this question, we go back to the motivation to look for the equilibrium solutions for the RDU agent.
Because the RDU preference is time-inconsistent, the current optimal strategy is not necessarily optimal in the future.
If the agent is unable to pre-commit such a time-inconsistent strategy, the agent should look for one that will not be deviated in the future. This leads to the concept of equilibrium solutions, which address this issue of time inconsistency. 
We assume that the agent will not deviate an SES once it is given at the initial time; otherwise we have to introduce another kind of equilibrium.
Under this assumption, it is natural for the agent to maximize the RDU of the initial time over all SESes. So we introduce the following definition.

\begin{defn} Let $\Ec$ denote all DSESes.  Let $\bm{\overbar{\pi}}\in\Ec$. We say that $\bm{\overbar{\pi}}$ is an optimal DSES if $J(0,x;\bm{\overbar{\pi}})\ge J(0,x;\bm{\pi})$ for all $\bm{\pi}\in\Ec$. 
\end{defn}

To our best knowledge, there is no existing work to discuss the optimal equilibria for the time-inconsistent control problems but three recent papers: \citet*{hz2019,hz2020mf} and \citet*{hw2021}, where the optimal equilibria for time-inconsistent stopping problems are discussed. 

The next lemma provides a formula of the RDU of a DSES. 
\begin{lem}\label{lma:rdu:form}
Let $\gamma=0$, $\bm{\mu}\ne0$ and Assumption \ref{assu:DistFun-t} hold.
Assume that $h\in C^2([0,T)\times(0,\infty))$, $h_{x}(\cdot,0)>0$ on $[0,T)$ and $\limsup_{t\uparrow T}\frac{h_{x}(t,0)}{\theta\sqrt{T-t}}<1$. Let $\bm{\pi}$ be the DSES given by \eqref{eq:pi-eq-t-log} with $Y$ being a positive solution to ODE  (\ref{eq:Y-ode-log-t}). Then
\begin{equation}\label{eq:J:form:log}
J(t,x;\boldsymbol{\pi}) = \ln x+r(T-t)+\int_{t}^{T}l\left(\sqrt{Y(s)};s\right)\dt s,\quad (t,x)\in[0,T)\times(0,\infty),
\end{equation}
where 
\[
l(z;s)\triangleq\frac{1}{2}\theta^{2}\frac{z}{z+h_{x}(s,0)}+zh_{tx}\left(s,0\right),\quad z\in[0,\infty).
\]
\end{lem}

\begin{proof} 
It follows from (\ref{eq:k-coe-t}) that 
\[
\bm{\mu}^{\top}\boldsymbol{\pi}(t)=\theta^{2}\left[1+\frac{h_{x}\left(t,0\right)}{\sqrt{\Pi(t)}}\right]^{-1}.
\]
Then similar to \eqref{eq:f:log}, we have
\begin{align*}
J(t,x;\boldsymbol{\pi})  =&\ln x+r(T-t)+\int_{t}^{T}\boldsymbol{\pi}^{\top}(s)\bm{\mu}\dt s-\frac{1}{2}\Pi(t)-\sqrt{\Pi(t)}h_{x}\left(t,0\right)\\
= & \ln x+r(T-t)+\int_{t}^{T}\theta^{2}\left[1+\frac{h_{x}\left(s,0\right)}{\sqrt{\Pi(s)}}\right]^{-1}\dt s+\frac{1}{2}\int_{t}^{T}\Pi'(s)\dt s\\
  & +\int_{t}^{T}\left[\sqrt{\Pi(s)}h_{tx}\left(s,0\right)+\frac{\Pi'(s)}{2\sqrt{\Pi(s)}}h_{x}\left(s,0\right)\right]\dt s\\
  = & \ln x+r(T-t)+\int_{t}^{T}\left\{ \theta^{2}\left[1+\frac{h_{x}\left(s,0\right)}{\sqrt{\Pi(s)}}\right]^{-1}+\sqrt{\Pi(s)}h_{tx}\left(s,0\right)\right\} \dt s\\
   & +\frac{1}{2}\int_{t}^{T}\Pi'(s)\left[1+\frac{1}{\sqrt{\Pi(s)}}h_{x}\left(s,0\right)\right]\dt s.
\end{align*}
Note that $\Pi=Y$ is the solution to  ODE \eqref{eq:Y-ode-log-t}. We have
\[
\Pi'(t)=Y'(t)=-\theta^{2}\left[\frac{\sqrt{Y(t)}}{\sqrt{Y(t)}+h_{x}(t,0)}\right]^{2},\quad t\in[0,T).
\]
Then 
\begin{align*}
J(t,x;\boldsymbol{\pi})  = & \ln x+r(T-t)+\int_{t}^{T}\left\{ \theta^{2}\left[1+\frac{h_{x}\left(s,0\right)}{\sqrt{Y(s)}}\right]^{-1}+\sqrt{Y(s)}h_{tx}\left(s,0\right)\right\} \\
   & -\frac{1}{2}\int_{t}^{T}\theta^{2}\left[\frac{\sqrt{Y(s)}}{\sqrt{Y(s)}+h_{x}(s,0)}\right]^{2}\left[1+\frac{1}{\sqrt{Y(s)}}h_{x}\left(s,0\right)\right]\dt s\\
  = & \ln x+r(T-t)+\int_{t}^{T}l\left(\sqrt{Y(s)};s\right)\dt s.
\end{align*}
\end{proof}

Obviously, the partial derivative of $l$
\[
l_{z}(z;s)=\frac{1}{2}\theta^{2}\frac{h_{x}(s,0)}{\left(z+h_{x}(s,0)\right)^{2}}+h_{tx}\left(s,0\right)
\]
is decreasing with respective to $z$.
Then we have the following proposition.

\begin{prop}
\label{prop:opt-eq}  
Let $\gamma=0$, $\bm{\mu}\ne0$ and Assumption \ref{assu:DistFun-t} hold.
Assume that $h\in C^2([0,T)\times(0,\infty))$, $h_{x}(\cdot,0)>0$ on $[0,T)$ and $\limsup_{t\uparrow T}\frac{h_{x}(t,0)}{\theta\sqrt{T-t}}<1$. Let $\overbar{Y}$ be the maximal solution to the ODE  (\ref{eq:Y-ode-log-t}).  Let $\bm{\overbar{\pi}}$ be given by 
\begin{equation}\label{eq:opt:barY}
\boldsymbol{\overbar{\pi}}(t)=\left(\bm{\sigma}\bm{\sigma}^{\top}\right)^{-1}\bm{\mu}\left[1+\frac{h_{x}\left(t,0\right)}{\sqrt{\overbar{Y}(t)}}\right]^{-1},\quad t\in[0,T).
\end{equation}
If $l_{z}\left(\sqrt{\overbar{Y}(s)};s\right)\geq0$ for all $s\in[0,T)$,
then $\boldsymbol{\overbar{\pi}}$ is an optimal DSES; moreover, it is uniformly optimal:
$J(t,x;\bm{\overbar{\pi}})\ge J(t,x;\bm{\pi})$ for all $(t,x)\in[0,T)\times(0,\infty)$ and $\bm{\pi}\in\Ec$.
\end{prop}

\begin{proof}
For all $s\in[0,T)$, by the monotonicity of $l_z(z;s)$, we have $l_{z}\left(z;s\right)\geq l_{z}\left(\sqrt{\overbar{Y}(s)}\right)\geq0$
for $z\in[0,\sqrt{\overbar{Y}(s)}]$. Then, for all $(t,x)\in[0,T)\times(0,\infty)$,
\begin{align*}
J(t,x;\boldsymbol{\overbar{\pi}})  = & \ln x+r(T-t)+\int_{t}^{T}l\left(\sqrt{\overbar{Y}(s)};s\right)\dt s\\
  \geq & \ln x+r(T-t)+\int_{t}^{T}l\left(\sqrt{Y(s)};s\right)\dt s
\end{align*}
for all positive solutions $Y$ to ODE  (\ref{eq:Y-ode-log-t}) as well as for $Y\equiv0$. Therefore,  by Lemma \ref{lma:rdu:form}, $\boldsymbol{\overbar{\pi}}$ is  uniformly optimal.
\end{proof}

For power utility, the RDU at time $t$ of a strategy $\bm{\pi}\in\Ec$ is 
\[
J(t,x;\boldsymbol{\pi})=\frac{1}{\gamma}x^{\gamma}\exp\left\{ \gamma\left[r(T-t)+\int_{t}^{T}\boldsymbol{\pi}^{\top}(s)\bm{\mu}\dt s-\frac{1}{2}\Pi(t)\right]\right\} h\left(t,-\gamma\sqrt{\Pi(t)}\right).
\]
Similar to Lemma \ref{lma:rdu:form}, under the conditions of Theorem \ref{thm:dete-eq-t-pstv}(ii), for a DSES $\bm{\pi}$  generated by \eqref{eq:pi-eq-t} with $Y$ being a positive solution to ODE \eqref{eq:Y-ode-t},  
one has 
\begin{equation}\label{eq:J:form:power}
J(t,x;\boldsymbol{\pi})  =  \frac{1}{\gamma}x^{\gamma}e^{\gamma r(T-t)}\exp\left\{ \int_{t}^{T}\left(\frac{\frac{\gamma}{2}\theta^{2}}{1+\frac{h_{x}\left(s,-\gamma\sqrt{Y(s)}\right)}{h\left(s,-\gamma\sqrt{Y(s)}\right)\sqrt{Y(s)}}}-\frac{h_{t}\left(s,-\gamma\sqrt{Y(s)}\right)}{h\left(s,-\gamma\sqrt{Y(s)}\right)}\right)\dt s\right\}.
\end{equation}
This formula in the power utility case is much more complicate than that one in the logarithmic utility case. Therefore, it is difficult,  for a power utility function and a general probability weighting function, to get an analogue of Proposition \ref{prop:opt-eq}.  We proceed with the discussion 
for a specific family of probability weighting function in the next example.

\begin{example}
\label{exam:ybar-new} Let $\bm{\mu}\ne0$, $\beta\in(0,\theta)$ and for any $t\in[0,T)$, 
$$w(t,p)=\Phi\left(\frac{1}{\lambda}\left(\Phi^{-1}(p)-\beta\sqrt{T-t}\right)\right),\quad p\in[0,1].$$
By the results in Example \ref{exm:w:Phi}, we have
\begin{align*}
h(t,x)=&e^{\beta\sqrt{T-t}x+\frac{\lambda^2}{2}x^2},\quad(t,x)\in[0,T)\times[0,\infty),\\
h_x(t,x)=&\left(\beta\sqrt{T-t}+\lambda^2x\right)h(t,x), \quad(t,x)\in[0,T)\times[0,\infty),\\
h_{t}(t,x)=&-\frac{x\beta}{2\sqrt{T-t}}h(t,x),\quad(t,x)\in[0,T)\times[0,\infty),\\ 
h_{tx}(t,0)=&-\frac{\beta}{2\sqrt{T-t}}, \quad t\in[0,T).
\end{align*}
Clearly, all of Assumptions \ref{assu:DistFun-t},  \ref{assu:h-t} and \ref{ass:zeta} are satisfied. 
For $\gamma\leq 0$,  ODE (\ref{eq:Y-ode-t})
reduces to
\begin{equation}
\begin{cases}
Y'(t)=-\theta^{2}\left(1-\gamma\lambda^{2}+\beta\frac{\sqrt{T-t}}{\sqrt{Y(t)}}\right)^{-2}, & t\in[0,T),\\
Y(T)=0.
\end{cases}\label{eq:Y-ode-t-example-p}
\end{equation}
For any positive solution $Y$ to (\ref{eq:Y-ode-t-example-p}), we
have 
\begin{align*}
Y'(t)&=-\theta^{2}\left(1-\gamma\lambda^{2}+\beta\frac{\sqrt{T-t}}{\sqrt{Y(t)}}\right)^{-2}\\
& \geq-\theta^{2}\left(2\sqrt{1-\gamma\lambda^{2}}\sqrt{\beta\frac{\sqrt{T-t}}{\sqrt{Y(t)}}}\right)^{-2}\\
 & =-\frac{\theta^{2}}{4\left(1-\gamma\lambda^{2}\right)\beta\sqrt{T-t}}\sqrt{Y(t)}
\end{align*}
and hence
\[
\frac{\dt\sqrt{Y(t)}}{\dt t}=\frac{Y'(t)}{2\sqrt{Y(t)}}\geq-\frac{\theta^{2}}{8\left(1-\gamma\lambda^{2}\right)\beta\sqrt{T-t}}.
\]
Then we have
\[
\sqrt{Y(t)}\leq\frac{\theta^{2}}{4\left(1-\gamma\lambda^{2}\right)\beta}\int_{t}^{T}\frac{\dt s}{2\sqrt{T-s}}=\frac{\theta^{2}}{4\left(1-\gamma\lambda^{2}\right)\beta}\sqrt{T-t}\quad\forall t\in[0,T).
\]
Therefore, the maximal positive solution $\overbar{Y}$ to (\ref{eq:Y-ode-t-example-p})
satisfies
\[
\overbar{Y}(t)\leq\left[\frac{\theta^{2}}{4\left(1-\gamma\lambda^{2}\right)\beta}\right]^{2}(T-t).
\]
In the specific case $\beta=\theta/2$,  $Y(t)=\left[\frac{\theta}{2\left(1-\gamma\lambda^{2}\right)}\right]^{2}(T-t)$
solves (\ref{eq:Y-ode-t-example-p}) and therefore, the maximal solution
\[
\overbar{Y}(t)=\left[\frac{\theta}{2\left(1-\gamma\lambda^{2}\right)}\right]^{2}(T-t),\quad t\in[0,T)
\]
and the corresponding DSES is  
\begin{equation*}
\boldsymbol{\overbar{\pi}}(t)=\frac{1}{2(1-\gamma\lambda^2)}\left(\bm{\sigma}\bm{\sigma}^{\top}\right)^{-1}\bm{\mu},\quad t\in[0,T).
\end{equation*}
\begin{description}
\item[(i)] Let $\gamma=0$. Then we have
\[
l_{z}(z;s)=\frac{1}{2}\theta^{2}\frac{\beta\sqrt{T-s}}{\left(z+\beta\sqrt{T-s}\right)^{2}}-\frac{\beta}{2\sqrt{T-s}}
\]
and
\begin{align*}
l_{z}\left(\sqrt{\overbar{Y}(s)};s\right)
&\geq l_{z}\left(\frac{\theta^{2}}{4\beta}\sqrt{T-s};s\right)\\
 & =\frac{1}{2}\theta^{2}\frac{\beta\sqrt{T-s}}{\left(\frac{\theta^{2}}{4\beta}\sqrt{T-s}+\beta\sqrt{T-s}\right)^{2}}-\frac{\beta}{2\sqrt{T-s}}\\
 & =\frac{1}{2}\left[\frac{\theta^{2}}{\left(\frac{\theta^{2}}{4\beta}+\beta\right)^{2}}-1\right]\frac{\beta}{\sqrt{T-s}}\\
 & \geq0,
\end{align*}
where the last inequality holds if and only if $\beta=\theta/2$.

Therefore, if $\beta=\theta/2$, then $l_{z}\left(\sqrt{\overbar{Y}(s)};s\right)=0$
and by Proposition \ref{prop:opt-eq} the strategy $\bm{\overbar{\pi}}$ 
is uniformly optimal.

\item[(ii)] Let $\gamma<0$. By \eqref{eq:J:form:power}, we have
\begin{align*}
J(t,x;\boldsymbol{\pi})=&\frac{1}{\gamma}x^{\gamma}e^{\gamma r(T-t)}\exp\left\{ \int_{t}^{T}\frac{\gamma}{2}\left(\frac{\theta^{2}}{1-\gamma\lambda^{2}+\beta\frac{\sqrt{T-s}}{\sqrt{\Pi(s)}}}-\beta\frac{\sqrt{\Pi(s)}}{\sqrt{T-s}}\right)\dt s\right\}\\
=&\frac{1}{\gamma}x^{\gamma}e^{\gamma r(T-t)}\exp\left\{ \int_{t}^{T}\frac{\gamma}{2}p\left(\sqrt{\Pi(s)};s\right)\dt s\right\}, 
\end{align*}
where
\[
p(z;s)\triangleq\frac{\theta^{2}z}{(1-\gamma\lambda^{2})z+\beta\sqrt{T-s}}-\beta\frac{z}{\sqrt{T-s}},\quad (s,z)\in[0,T)\times[0,\infty).
\]
Clearly, for every $s\in[0,T)$,
\[
p_{z}(z;s)=\frac{\theta^{2}\beta\sqrt{T-s}}{\left((1-\gamma\lambda^{2})z+\beta\sqrt{T-s}\right)^{2}}-\frac{\beta}{\sqrt{T-s}}
\]
is decreasing w.r.t. $z$. Then
\begin{align*}
p_{z}\left(\sqrt{\overbar{Y}(s)};s\right)&\geq p_{z}\left(\frac{\theta^{2}}{4\left(1-\gamma\lambda^{2}\right)\beta}\sqrt{T-s};s\right)\\
 & =\frac{\theta^{2}\beta\sqrt{T-s}}{\left((1-\gamma\lambda^{2})\frac{\theta^{2}}{4\left(1-\gamma\lambda^{2}\right)\beta}\sqrt{T-s}+\beta\sqrt{T-s}\right)^{2}}-\frac{\beta}{\sqrt{T-s}}\\
 & =\left[\frac{\theta^{2}}{\left(\frac{\theta^{2}}{4\beta}+\beta\right)^{2}}-1\right]\frac{\beta}{\sqrt{T-s}}\\
 & \geq0,
\end{align*}
where the last inequality holds if and only if $\beta=\theta/2$.

Therefore, if $\beta=\theta/2$, then
$p_{z}\left(\sqrt{\overbar{Y}(s)};s\right)=0$ and the strategy $\bm{\overbar{\pi}}$ 
is uniformly optimal.
\end{description}
Overall, when $\beta=\theta/2$, the equilibrium strategy $\bm{\overbar{\pi}}$ generated by the maximal positive solution $\overbar{Y}$ is uniformly optimal. \eed
\end{example}

In general, however, we have to search the optimal DSES numerically. It can be carried out as follows.
\begin{itemize}
\item First, determine the value of $\eta^*$ as well as the maximal positive solution $\overbar{Y}$ by the method provided in Remark \ref{rem:approximate eta}. That is, solve the backward ODE \eqref{eq:bode:Y;epsilon}
for sufficiently small $\varepsilon>0$ to get approximations of $\overbar{Y}$ and $\eta^*=\overbar{Y}(0)$.
\item Second, search for $\eta\in[0,\eta^*]$ to maximize the RDU $J(0,x,\bm{\pi})$, which, by \eqref{eq:J:form:log} or \eqref{eq:J:form:power}, is a function of $\eta$ since $\bm{\pi}\in\Ec$ if and only if it is generated by \eqref{eq:pi-eq-t} with $Y$ being the unique solution of the forward ODE \eqref{eq:Y-ode-t-forward} with the initial condition $Y(0)= \eta\in[0,\eta*]$.
\end{itemize}
 Now we conduct a numerical analysis with the following values of the parameters: $d=n=1$,
$x=1$, $T=10$, $r=0$, $\mu=0.05$, $\sigma=0.2$, $\gamma=-2$.
Figure \ref{fig:rdu0-pi} depicts the RDU of the initial time
as a function of $\eta\in[0,\eta^*]$ (note that $\eta^*$ depends on $\lambda$ and $\beta$) and the corresponding optimal equilibrium strategies for
 various $\lambda>0$ and $\beta\in(0,\theta)$. It can be seen that the optimal equilibria is attained at some $\eta\in(0,\eta^*)$ and the probability distortion does affect the solution significantly. The Merton solution is $\bm{\pi}_{m}=\theta/(\sigma(1-\gamma))=41.67\%$, whereas the optimal equilibrium solutions are from 2.37\% to 4.63\%, which are much smaller than the Merton solution. In particular, in the case $\lambda=1$, when $t$ is approaching $T$, the probability weighting function $w(t,p)\to p$, but 
 the optimal equilibrium solutions are still much smaller than the Merton solution. On the other hand, the optimal equilibrium solution is decreasing in $\lambda$ and $\beta$.

\begin{center}
\begin{figure}[h!]
\begin{centering}
\subfloat[$\lambda=1.5$]{\begin{centering}
\includegraphics[scale=0.45]{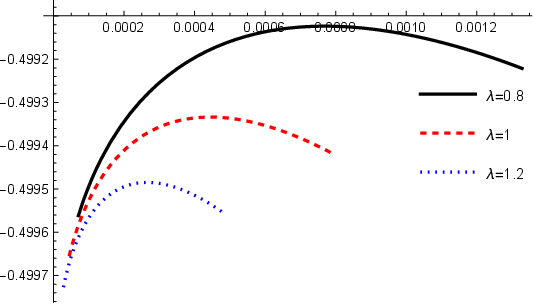}\includegraphics[scale=0.45]{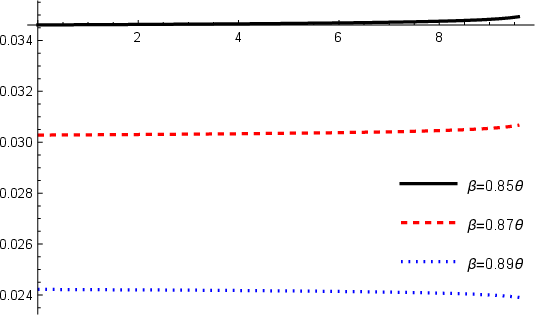}
\par\end{centering}
}
\par\end{centering}

\begin{centering}
\subfloat[$\beta=0.92\theta$]{\begin{centering}
\includegraphics[scale=0.45]{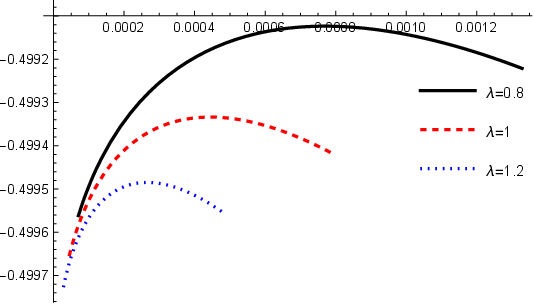}\includegraphics[scale=0.45]{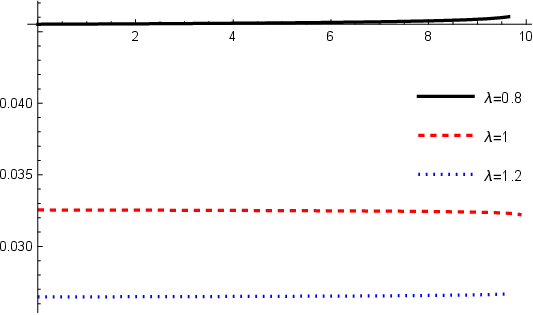}
\par\end{centering}
}
\par\end{centering}
\caption{RDU at time $0$ (left panel) and the optimal equilibrium strategy (right panel)}\label{fig:rdu0-pi} 
\end{figure}
\par\end{center}

\section{Concluding Remarks}

This paper addresses the problem of dynamic portfolio selection within
the framework of RDU theory. The challenge of this problem is that the optimization
problem is time-inconsistent and the classical method such as dynamic
programming does not apply. Considering an incomplete financial market
with constant coefficients and CRRA utilities, we identify DSESes. 
In the case of time-invariant probability weighting function, we conduct a thorough analysis that covers all cases: there is up to one DSES in any case and, if the DSES exits, it is worked out in closed form.  
In the case of time-variant probability weighting functions, the non-zero DSESes are characterized by the positive solutions
to a nonlinear singular ODE, which is investigated under various assumptions. 
In particular, if the positive solutions are not unique (and so are the DSESes), we provide a method to find out all the positive solutions to this singular ODE and identify the optimal equilibria among all the DSESes. 
It is worth mentioning that our method of characterizing all the positive solutions also applies to
the singular ODE in \citet*{hjz2021}. Finally, it should be noted that investigating non-deterministic equilibrium strategies for portfolio selection under RDU is very  challenging. We even do not know whether there is a non-deterministic equilibrium strategy.

\newpage
\begin{appendices}

\section{Function $h$ \protect\label{proof:lem:h}}

Recall that $\Phi$ is the standard normal distribution function.
Then by \eqref{eq:h},
\begin{equation}
h(x)=\Ebf\left[e^{-x\xi}w^{\prime}(1-\Phi(\xi))\right]=\Ebf\left[e^{x\xi}w'(\Phi(\xi))\right],\quad x\in\mathbb{R},\label{eq:fn-h}
\end{equation}
which coincides the time-varying version of the definition in \citet*{hjz2021}.
The following Lemma \ref{lem:h} extends
some results of \citet*[Lemma 2.7]{hjz2021} from the half line $[0,\infty)$ to the whole line $\Rbb$.
\begin{lem}
\label{lem:h} Under Assumption \ref{assu:DistFun}, the function
$h$ has the following properties: 
\begin{description}
\item[(i)] $h(0)=1$ and $h(x)<\infty$ $\forall x\in\mathbb{R}$; 
\item[(ii)] $h$ is $C^{\infty}$ on $\mathbb{R}$ and its $k$-th derivative
\begin{equation}
h^{(k)}(x)=\int_{0}^{1}\left[-Q_{\xi}(p)\right]^{k}e^{-xQ_{\xi}(p)}\dt\overbar{w}(p)=\int_{0}^{1}\left[Q_{\xi}(p)\right]^{k}e^{xQ_{\xi}(p)}\dt w(p),\quad x\in\mathbb{R},\,k\geq0;\label{eq:h^(n)}
\end{equation}
\item[(iii)] $h^{(k)}$ is strictly positive and strictly convex on $\mathbb{R}$ for all even $k\geq0$,
and strictly increasing for all odd $k\geq1$. 
\end{description}
\end{lem}

\begin{proof}
Assertion (i) is a consequence of \citet*[Lemmas 2.6 and 2.7]{hjz2021}. Moreover,
assertion (iii) is a consequence of assertion (ii). Therefore, it is left to show assertion (ii).
For any probability weighting function $w$, let 
\[
g_{w}(x)\triangleq\Ebf\left[e^{-x\xi}w^{\prime}(1-\Phi(\xi))\right]=\Ebf\left[e^{x\xi}w'(\Phi(\xi))\right],\quad x\in\mathbb{R}.
\]
Then, by \eqref{eq:fn-h}, 
\begin{align*}
h(x)=\Ebf\left[e^{-x\xi}w'(1-\Phi(\xi))\right]=\Ebf\left[e^{-x\xi}\overbar{w}'(\Phi(\xi))\right]=g_{\overbar{w}}(-x),\quad x\in\Rbb.
\end{align*}
Obviously, $\overbar{w}$ satisfies Assumption \ref{assu:DistFun}
as well. Then by \citet*[Lemma 2.7]{hjz2021}, 
\[
h^{(k)}(x)=g_{w}^{(k)}(x)=\Ebf\left[\xi^{k}e^{x\xi}w'(\Phi(\xi))\right],\quad x\ge0
\]
and hence 
\[
h^{(k)}(x)=(-1)^{k}g_{\overbar{w}}^{(k)}(-x)=(-1)^{k}\Ebf\left[\xi^{k}e^{-x\xi}\overbar{w}'(\Phi(\xi))\right],\quad x\le0.
\]
Moreover, 
\[
h^{(k)}(0+)=\Ebf\left[\xi^{k}w'(\Phi(\xi))\right]=\Ebf\left[(-\xi)^{k}\overbar{w}'(\Phi(\xi))\right]=h^{(k)}(0-).
\]
Consequently, assertion (ii) follows.
\end{proof}

\section{ODE \eqref{eq:Y-ode}\protect\label{sec:ode}}

The following proposition presents some results on the positive solutions to ODE \eqref{eq:Y-ode}.

\begin{prop}
\label{prop:exist-Y}Suppose Assumption \ref{assu:DistFun} holds. 
\begin{description}
\item[(i)] If ODE (\ref{eq:Y-ode}) admits a positive solution, then $\bm{\mu}\ne\bm{0}$ and $h'(0)=0$; 
\item[(ii)] If $\bm{\mu}\ne\bm{0}$, $h'(0)=0$, 
and $\mathcal{G}\left(y_{1}\right)>\theta^{2}T$ (where $\Gc$ is given by \eqref{eq:Gc}),
then 
\begin{equation}
Y(t)=\mathcal{G}^{-1}\left(\theta^{2}(T-t)\right),\quad t\in[0,T]\label{eq:Y}
\end{equation}
is the unique positive solution to ODE (\ref{eq:Y-ode}) and $Y(t)\in\mbb{Y}$
for $t\in[0,T)$.
\end{description}
\end{prop}

\begin{proof}
\begin{enumerate}
\item[(i)]  Let $Y$ be a positive solution to ODE (\ref{eq:Y-ode}). Obviously, we have $\bm{\mu}\ne\bm{0}$. Let $y$ be given by $y(t)=Y(T-t)$, $t\in[0,T]$. Then $y(t)>0$ for all $t\in(0,T]$ and $y$ satisfies the following ODE:
\begin{equation}
\begin{cases}
y'(t)=\theta^{2}\left[1+\frac{h'\left(-\gamma\sqrt{y(t)}\right)}{h\left(-\gamma\sqrt{y(t)}\right)\sqrt{y(t)}}\right]^{-2}, & t\in(0,T],\\
y(0)=0.
\end{cases}\label{eq:y-ode}
\end{equation}
It is left to show $h'(0)=0$.
Suppose on the contrary that $h'(0)\neq0$.
Then there exists $\varepsilon>0$ and $\delta\in(0,T)$ such that
\[
\left|h\left(-\gamma\sqrt{y(t)}\right)\sqrt{y(t)}+h'\left(-\gamma\sqrt{y(t)}\right)\right|^{2}>\varepsilon\quad\forall\,t\in[0,\delta].
\]
Therefore, for any given $t\in[0,\delta]$, 
\[
\int_{0}^{t}\theta^{2}\left[\frac{h\left(-\gamma\sqrt{y(s)}\right)}{h\left(-\gamma\sqrt{y(s)}\right)\sqrt{y(s)}+h'\left(-\gamma\sqrt{y(s)}\right)}\right]^{2}\dt s<\infty.
\]
If follows from Gronwall's inequality that 
\[
y(t)\leq\exp\left\{ \int_{0}^{t}\theta^{2}\left[\frac{h\left(-\gamma\sqrt{y(s)}\right)}{h\left(-\gamma\sqrt{y(s)}\right)\sqrt{y(s)}+h'\left(-\gamma\sqrt{y(s)}\right)}\right]^{2}\dt s\right\} y(0)=0,\quad t\in[0,\delta],
\]
which is impossible since $y(t)>0$ for all $t\in(0,T]$.
\item[(ii)] Note that  $\mathcal{G}\left(y_{1}\right)>\theta^{2}T$ implies $y_1>0$. Obviously, $Y$ is a positive solution to ODE (\ref{eq:Y-ode}) and $Y(t)\in(0,y_{1})\subseteq\mbb{Y}$ for all $t\in[0,T)$.
Assume that  $\overbar{Y}$ is a positive solutions to (\ref{eq:Y-ode}). Then 
\begin{align*}
\theta^{2}(T-t)= &- \int_t^T\left[\frac{h\left(-\gamma\sqrt{\overbar{Y}(s)}\right)}{h\left(-\gamma\sqrt{\overbar{Y}(s)}\right)\sqrt{\overbar{Y}(s)}+h'\left(-\gamma\sqrt{\overbar{Y}(s)}\right)}\right]^{-2}\frac{\overbar{Y}'(s)}{\overbar{Y}(s)}\dt s\\
= & \int^{\overbar Y(t)}_{0}\left[\frac{h\left(-\gamma\sqrt{z}\right)}{h\left(-\gamma\sqrt{z}\right)\sqrt{z}+h'\left(-\gamma\sqrt{z}\right)}\right]^{-2}\frac{1}{z}\dt z\\
= & \mathcal{G}(\overbar{Y}(t)),
\end{align*}
which implies that $\overbar{Y}(t)=\mathcal{G}^{-1}\left(\theta^{2}(T-t)\right)=Y(t)$.
Thus, we have the uniqueness of the positive solution.\qedhere
\end{enumerate}
\end{proof}

\section{ODE \eqref{eq:Y-ode-t} with $\gamma<0$\protect\label{sec:ode-t}}

ODE (\ref{eq:Y-ode-t}) will be studied under various assumptions.

\subsection{Zero Solution}
\begin{prop}
\label{prop:exist-Y-t-0} Let $\gamma<0$. Under Assumption \ref{assu:DistFun-t}, if  $h$ and $h_x$ are continuous on $[0,T)\times[0,\infty)$ and $\inf_{t\in[0,T)}h_{x}(t,0)>0$, then $0$ is the unique solution to ODE (\ref{eq:Y-ode-t}). 
\end{prop}

\begin{proof} 
Obviously, $0$ solves ODE (\ref{eq:Y-ode-t}) since $m(t,0)=0$ for all $t\in[0,T)$. We are going to prove the uniqueness. Let $Y$ be a solution to ODE (\ref{eq:Y-ode-t}).
Let $y(t)=Y(T-t)$, $t\in[0,T]$. Then $y\in AC([0,T])\cap C^1((0,T])$ is an increasing function that solves the following ODE: 
\begin{equation*}
\begin{cases}
y'(t)=\theta^{2}m^{2}\left(T-t,\sqrt{y(t)}\right), & t\in(0,T],\\
y(0)=0.
\end{cases}
\end{equation*}
For any $t\in[0,T]$, by the convexity of $h(T-t,\cdot)$ (Lemma \ref{lem:h}(iii)), 
\begin{align*}
y'(t)=&\theta^{2}\left[\frac{h\left(T-t,-\gamma\sqrt{y(t)}\right)\sqrt{y(t)}} {h\left(T-t,-\gamma\sqrt{y(t)}\right)\sqrt{y(t)}+h_x\left(T-t,-\gamma\sqrt{y(t)}\right)}
\right]^2 \\
 \leq & \theta^{2}\left[\frac{h\left(T-t,-\gamma\sqrt{y(t)}\right)}{h_{x}\left(T-t,-\gamma\sqrt{y(t)}\right)}\right]^{2}y(t)\\
  \leq & \theta^{2}\left[\frac{h\left(T-t,0\right)+h_{x}\left(T-t,-\gamma\sqrt{y(t)}\right)\left(-\gamma\sqrt{y(t)}\right)}{h_{x}\left(T-t,-\gamma\sqrt{y(t)}\right)}\right]^{2}y(t)\\
  \leq & \theta^{2}\left[\frac{1}{\inf_{t\in[0,T)}h_{x}\left(t,0\right)}-\gamma\sqrt{y(T)}\right]^{2}y(t).
\end{align*}
By Gronwall's inequality, $y(t)\le 0$ for all $t\in[0,T]$. Therefore, $y=0$ on $[0,T]$. The uniqueness is proved.
\end{proof}

\subsection{Existence of a Positive Solution}

\begin{prop}
\label{prop:exist-Y-t} Let $\gamma<0$ and $\bm{\mu}\ne0$. Under Assumptions \ref{assu:DistFun-t}
and \ref{assu:h-t}, ODE (\ref{eq:Y-ode-t}) has a positive solution. 
\end{prop}

\begin{proof}
The conclusion is a consequence of a combination of Lemmas \ref{lma:positive1} and \ref{lma:positive2} below.
\end{proof}

The next lemma can be proved by mimicking the proof of  \citet*[Theorem 3.1]{ao2004}.

\begin{lem}
\label{lem:singular-ode} 
Let $\tau>0$. Consider the ODE 
\begin{equation}
\begin{cases}
y'(t)=l(t,y(t)),\quad t\in(0,\tau),\\
y(0)=0.
\end{cases}\label{eq:singular-ODE}
\end{equation}
Suppose the following conditions
are satisfied: 
\begin{description}
\item[(i)] $l :(0,\tau]\times(0,\infty)\rightarrow\mathbb{R}$ is continuous and bounded; 

\item[(ii)] There exists a function $\alpha\in C([0,\tau])\cap C^{1}((0,\tau])$ such that $\alpha(0)=0$, 
$\alpha(t)>0$ for $t\in(0,\tau]$ and $l(t,\alpha(t))\geq\alpha'(t)$ for $t\in(0,\tau)$.
\end{description}
Then (\ref{eq:singular-ODE}) admits a solution $y\in C([0,\tau])\cap C^{1}((0,\tau])$
with $y(t)\geq\alpha(t)$ for $t\in[0,\tau]$. 
\end{lem}

\begin{lem}\label{lma:positive1}
Let $\gamma<0$ and $\bm{\mu}\ne0$. Under Assumptions \ref{assu:DistFun-t} and \ref{assu:h-t}, there exists some $\tau\in(0,T)$ such that the following
ODE 
\begin{equation}
\begin{cases}
y'(t)=\theta^{2}m^{2}\left(T-t,\sqrt{y(t)}\right), & t\in(0,\tau),\\
y(0)=0
\end{cases}\label{eq:y-ode-t-(0,delta)}
\end{equation}
has a solution $y\in C([0,\tau])\cap C^{1}((0,\tau])$ such that $y(t)>0$ for all $t\in(0,\tau]$. 
\end{lem}

\begin{proof} Let $l(t,y)=\theta^{2} m^{2}(T-t,\sqrt{y})$, $(t,y)\in(0,T]\times(0,\infty)$.

First, by conditions (i)--(ii) of Assumption \ref{assu:h-t}, condition (i) of Lemma \ref{lem:singular-ode} is
satisfied for every $\tau\in(0,T)$.

Second, we verify condition (ii)  of Lemma \ref{lem:singular-ode} for a sufficiently small $\tau\in(0,T)$. Let
\[
\alpha(t)=kt,\quad t\in[0,T],
\]
where $k>0$ is a sufficiently small constant to be determined. One needs to show that, for all sufficiently small $t\in(0,T)$, 
\[
\theta^{2}m^{2}\left(T-t,\sqrt{\alpha(t)}\right)=\left(\frac{1}{\frac{1}{\theta}\left(1+\frac{h_{x}\left(T-t,-\gamma\sqrt{kt}\right)}{h\left(T-t,-\gamma\sqrt{kt}\right)\sqrt{kt}}\right)}\right)^{2}\geq k,
\]
which is equivalent to 
\begin{equation}
\frac{\sqrt{k}}{\theta}+\frac{h_{x}\left(T-t,-\gamma\sqrt{kt}\right)}{\theta h\left(T-t,-\gamma\sqrt{kt}\right)\sqrt{t}}\leq1.\label{eq:a-t-1}
\end{equation}
By condition (ii) of Assumption \ref{assu:h-t}, we have $h(T-t,x)\ge1$ for all $(t,x)\in(0,T]\times[0,\infty)$. Then, by the mean value theorem and by conditions (iii)--(iv) of Assumption \ref{assu:h-t}, there exists $z(t)\in(0,-\gamma\sqrt{kt})$ such that
\begin{align*}
& \limsup_{t\downarrow0}\left(\frac{\sqrt{k}}{\theta}+\frac{h_{x}\left(T-t,-\gamma\sqrt{kt}\right)}{\theta h\left(T-t,-\gamma\sqrt{kt}\right)\sqrt{t}}\right)\\
 \le &  \frac{\sqrt{k}}{\theta}+\limsup_{t\downarrow0} \frac{h_{x}\left(T-t,-\gamma\sqrt{kt}\right)}{\theta\sqrt{t}}\\
 =& \frac{\sqrt{k}}{\theta}+\limsup_{t\downarrow0} \frac{h_{x}\left(T-t,0\right)+h_{xx}\left(T-t,z(t)\right)(-\gamma\sqrt{kt})}{\theta\sqrt{t}}\\
 \leq & \frac{\sqrt{k}}{\theta}-{\gamma\sqrt{k}\over\theta}M+\limsup_{t\downarrow0}{h_x(T-t,0)\over \theta\sqrt{t}}\\
 <  & 1
\end{align*}
for all sufficiently small $k>0$,
where $M=\limsup_{t\downarrow0,x\downarrow0}h_{xx}(T-t,x)<\infty$.
Therefore, there exists $k>0$ and $\tau\in(0,T)$ such that (\ref{eq:a-t-1})
holds for all $t\in(0,\tau)$.

Finally, by Lemma \ref{lem:singular-ode}, ODE \eqref{eq:y-ode-t-(0,delta)} has a solution $y\in C([0,\tau])\cap C^{1}((0,\tau])$  such that $y(t)\ge \alpha(t)>0$
for all $t\in(0,\tau]$. 
\end{proof}

\begin{lem}\label{lma:positive2}
Let $\gamma<0$ and $\bm{\mu}\ne0$. Under Assumptions \ref{assu:DistFun-t} and \ref{assu:h-t}, for any $\tau\in(0,T)$ and $\eta>0$, the ODE 
\begin{equation*}
\begin{cases}
y'(t)=\theta^{2}m^{2}\left(T-t,\sqrt{y(t)}\right), &t\in(\tau,T],\\
y(\tau)=\eta
\end{cases}  
\end{equation*}
admits a solution $y\in C([\tau,T])\cap C^{1}((\tau,T])$ such that $y(t)\ge\eta>0$ for all $t\in[\tau,T]$. 
\end{lem}

\begin{proof}
For any $(t,x)\in(0,T]\times(0,\infty)$, $0<m^{2}\left(T-t,x\right)\leq1$. Moreover, $m(T-t,x)$ is continuous on $(0,T]\times(0,\infty)$. Then
the result follows from the Peano existence theorem.
\end{proof}

\subsection{Characterization of Positive Solutions}\label{app:FB}

In this part, we provide a ``forward” method to find out all positive solutions to the backward ODE \eqref{eq:Y-ode-t}.
It turns out that all positive solutions of the backward ODE can be determined by solving a class of forward ODEs.

\begin{lem}
\label{lem:multiple-1} Let $\gamma<0$ and $\bm{\mu}\ne0$. Under Assumptions \ref{assu:DistFun-t}, \ref{assu:h-t}(i) and \ref{ass:zeta}(i,ii),
for any $\eta>0$, the forward ODE \eqref{eq:Y-ode-t-forward}
admits a unique solution $Y\in C([0,T])\cap C^1([0,T))$. 
\end{lem}

\begin{proof} Let $\eta>0$ be fixed. 
Let $\{\delta_j,j\ge1\}\subset(0,T)$ be an increasing sequence such that $\lim_{j\to\infty}\delta_j=T$. For every $j\ge1$,   consider the equation 
\begin{equation}
\begin{cases}
Y'(t)=-\theta^{2}m^{2}\left(t,\sqrt{Y(t)}\right), & t\in[0,\delta_j),\\
Y_j(0)=\eta.
\end{cases}\label{eq:y-ode-t-backward-delta}
\end{equation}
It is easy to see that 
${1\over m(t,x)}=1+{h_x(t,-\gamma x)\over h(t,-\gamma x)x}$ for all $(t,x)\in[0,T)\times(0,\infty)$. Differentiating both sides yields
\begin{equation}
\begin{split}
m_{x}(t,x)=m^{2}(t,x)\left\{ \frac{\gamma xh_{xx}\left(t,-\gamma x\right)+h_{x}\left(t,-\gamma x\right)}{h\left(t,-\gamma x\right)x^{2}}-\gamma x\left[\frac{h_{x}\left(t,-\gamma x\right)}{h\left(t,-\gamma x\right)x}\right]^{2}\right\},&\\
 (t,x)\in[0,T)\times(0,\infty).&
 \label{eq:m_x}
\end{split}
\end{equation}
Then, for all $(t,x)\in[0,T)\times(0,\infty)$,
\begin{align*}
\frac{d(m^{2}(t,x))}{dx} 
=& 2m(t,x)m_x(t,x)\\
= & 2m^{3}(t,x)\left\{ \frac{\gamma xh_{xx}\left(t,-\gamma x\right)+h_{x}\left(t,-\gamma x\right)}{h\left(t,-\gamma x\right)x^{2}}-\gamma x\left[\frac{h_{x}\left(t,-\gamma x\right)}{h\left(t,-\gamma x\right)x}\right]^{2}\right\} \\
= & 2\left(\frac{h(t,-\gamma x)x}{h(t,-\gamma x)x+h_{x}(t,-\gamma x)}\right)^{3}\\
   & \times\left\{ \frac{\gamma xh_{xx}\left(t,-\gamma x\right)+h_{x}\left(t,-\gamma x\right)}{h\left(t,-\gamma x\right)x^{2}}-\gamma x\left[\frac{h_{x}\left(t,-\gamma x\right)}{h\left(t,-\gamma x\right)x}\right]^{2}\right\} \\
  = & 2\left(\frac{h(t,-\gamma x)}{h(t,-\gamma x)x+h_{x}(t,-\gamma x)}\right)^{3}x\\
   & \times\left\{ \frac{\gamma xh_{xx}\left(t,-\gamma x\right)+h_{x}\left(t,-\gamma x\right)}{h\left(t,-\gamma x\right)}-\gamma x\left[\frac{h_{x}\left(t,-\gamma x\right)}{h\left(t,-\gamma x\right)}\right]^{2}\right\}.
\end{align*}
Thus, for all $(t,x)\in[0,T)\times(0,\infty)$,
\begin{align*}
\frac{d(m^{2}(t,\sqrt{x}))}{dx} = & \left(\frac{h(t,-\gamma\sqrt{x})}{h(t,-\gamma \sqrt{x})\sqrt{x}+h_{x}(t,-\gamma\sqrt{x})}\right)^{3}\\
  & \times\left\{ \frac{\gamma\sqrt{x}h_{xx}\left(t,-\gamma\sqrt{x}\right)+h_{x}\left(t,-\gamma\sqrt{x}\right)}{h\left(t,-\gamma\sqrt{x}\right)}-\gamma\sqrt{x}\left[\frac{h_{x}\left(t,-\gamma\sqrt{x}\right)}{h\left(t,-\gamma\sqrt{x}\right)}\right]^{2}\right\} .
\end{align*}
By Assumption \ref{ass:zeta}(i), $h_x(t,x)\ge h_x(t,0)>0$ and hence $h(t,x)\ge h(t,0)=1$ for all $(t,x)\in[0,T)\times[0,\infty)$. Then, by Assumptions \ref{assu:h-t}(i) and \ref{ass:zeta}(ii),
,
for every $j\ge1$, there exists a constant
$C_j>0$ such that 
\begin{equation}\label{ineq:m2C}
\left|\frac{d(m^{2}(t,\sqrt{x}))}{dx}\right|<C_j\quad\forall t\in[0,\delta_j]\text{ and }x\in[0,\eta],
\end{equation}
which implies the desired Lipschitz condition.  
Therefore, for each $j\ge1$, ODE (\ref{eq:y-ode-t-backward-delta})
admits a unique solution $Y_j\in C([0,\delta_j])\cap C^1([0,\delta_j))$, which is decreasing and non-negative on $[0,\delta_j]$. 
Let $Y(t)=Y_j(t)$ for $j\ge1$ and $t\in[0,\delta_j]$ and $Y(T)=\lim_{j\to\infty}Y_j(\delta_j)$. Then 
$Y$ is well defined,  $Y\in C([0,T])\cap C^1([0,T))$ and $Y$ is the unique solution to ODE \eqref{eq:Y-ode-t-forward}.
\end{proof}

\begin{rem}\label{rem:hat-eps}
By Assumption \ref{ass:zeta}(i),  $m(t,x)\in(0,1)$ for all $(t,x)\in[0,T)\times(0,\infty)$.
Let $\overhat Y$ be the unique solution to ODE \eqref{eq:Y-ode-t-forward} with $\eta=\theta^2T$. Then we have $\overhat\varepsilon\trieq\overhat Y(T)>0$.
\end{rem}

\begin{lem}
\label{lem:multiple-2} Let $\gamma<0$ and $\bm{\mu}\ne0$. Under Assumptions \ref{assu:DistFun-t},
\ref{assu:h-t}(i) and \ref{ass:zeta}(i,ii), if $Y_0$ is  a positive
solution to the backward ODE (\ref{eq:Y-ode-t}), then for any $\eta\in(0,Y_{0}(0)]$, the forward ODE (\ref{eq:Y-ode-t-forward})
admits a unique solution and the solution is also a positive solution to the backward ODE  (\ref{eq:Y-ode-t}). 
\end{lem}

\begin{proof}
Let $\eta\in(0,Y_{0}(0)]$.
By Lemma \ref{lem:multiple-1}, ODE (\ref{eq:Y-ode-t-forward}) admits 
a unique solution $Y$, which is obviously decreasing. It is left to show that $Y$ is also a positive solution to ODE  (\ref{eq:Y-ode-t}). Let 
\[
t_{0}=\sup\{t\in[0,T)\,:\,Y(t)>0\}.
\]
Obviously, $0<t_{0}\le T$. Suppose $0<t_{0}<T$. Then
$Y$ satisfies 
\begin{equation}
\begin{cases}
Y'(t)=-\theta^{2}m^{2}\left(t,\sqrt{Y(t)}\right), & t\in[0,t_{0}),\\
Y(t_{0})=0.
\end{cases}\label{eq:y-ode-t0}
\end{equation}
It follows from Proposition \ref{prop:exist-Y-t-0} that $0$ is the unique solution to (\ref{eq:y-ode-t0}). Then we have $Y=0$ on $[0,t_0)$,  which contradicts $Y(0)=\eta>0$. Therefore, $t_{0}=T$, which implies that $Y(T)\geq0$
and $Y(t)>0$ for $t\in[0,T)$. On the other hand, 
by \eqref{ineq:m2C} and by $\eta\leq Y_{0}(0)$, applying the comparison theorem to the forward ODEs satisfied by $Y$ and $Y_0$ on $[0,\delta]$ yields 
$Y\leq Y_{0}$ on $[0,\delta]$ for all $\delta\in(0,T)$. Hence, $Y(T)\leq Y_{0}(T)=0$.
Therefore, $Y(T)=0$ and $Y$ is a positive solution to ODE  (\ref{eq:Y-ode-t}). 
\end{proof}

 Recall that $\eta^*$ is given by \eqref{eq:eta*}. 

\begin{prop}
\label{prop:Y_M} Let $\gamma<0$ and $\bm{\mu}\ne0$. Under Assumptions \ref{assu:DistFun-t}, \ref{assu:h-t} and \ref{ass:zeta}, 
the maximal positive solution to the backward ODE (\ref{eq:Y-ode-t}) exists.  
\end{prop}

\begin{proof}
Let $\{\varepsilon_j,j\ge1\}\subset(0,\infty)$ be a strictly decreasing sequence such that $\lim_{j\to\infty}\varepsilon_j=0$.
For every $j\ge1$,   by Assumption \ref{ass:zeta}(iii), in a similar way to \eqref{ineq:m2C}, there exists $C_j>0$ such that
$$\left|\frac{d(m^{2}(t,\sqrt{x}))}{dx}\right|<C_j\quad\forall t\in[0,T)\text{ and }x\in[\varepsilon_j,\varepsilon_j+\theta^2 T].$$
Then the following backward ODE
\begin{equation}
\begin{cases}
Y'_j(t)=-\theta^{2}m^{2}\left(t,\sqrt{Y_j(t)}\right), & t\in[0,T),\\
Y_j(T)=\varepsilon_j 
\end{cases}\label{eq:Y-ode-t-epj}
\end{equation}
admits a unique solution $Y_{j}$.

Let $Y$ be a positive solution to the backward ODE (\ref{eq:Y-ode-t}). By the continuity of $Y_j$ and $Y$, there exists $\delta\in(0,T)$ such that $Y_j(t)> Y(t)$ for all $t\in[\delta,T]$. Recalling (\ref{ineq:m2C}) and applying the comparison theorem to the backward ODEs satisfied by $Y_j$ and $Y$ on $[0,\delta]$ yield that $Y_j> Y$ on $[0,\delta]$. Therefore, $Y_j>Y$ on $[0,T]$.  Similarly, $Y_j>Y_{j+1}$ on $[0,T]$ for all $j\ge1$ and $\overhat Y>Y_{j}$ on $[0,T]$ if $\overhat\varepsilon>\varepsilon_j$, where $\overhat Y$ and $\overhat\varepsilon$ are given in Remark \ref{rem:hat-eps}.
 
Let $\overbar{\eta}=\inf\{Y_j(0)\,:\,j\geq 1\}$. Then $\theta^2T=\overhat Y(0)\ge\overbar{\eta}\geq \eta^*>0$.
Let us consider the following forward ODE
\begin{equation}
\begin{cases}
\overbar{Y}'(t)=-\theta^{2}m^{2}\left(t,\sqrt{\overbar {Y}(t)}\right), & t\in[0,T),\\
\overbar{Y}(0)=\overbar{\eta}.
\end{cases}\label{eq:Y-ode-t-bareta}
\end{equation}
By Lemma \ref{lem:multiple-1}, ODE (\ref{eq:Y-ode-t-bareta}) admits a unique solution $\overbar{Y}$. For any positive solution $Y$ to ODE (\ref{eq:Y-ode-t}), we have $\overbar\eta\ge\eta^*\ge Y(0)$. Then,
by (\ref{ineq:m2C}), applying the comparison theorem to the forward ODEs satisfied by $\overbar{Y}$ and $Y$ on $[0,\delta]$ yields that $\overbar{Y}\geq Y$ on $[0,\delta]$ for any $\delta\in(0,T)$. Letting $\delta\rightarrow T$ yields that $\overbar{Y}\ge Y$ on $[0,T]$ and $\overbar{Y}>0$ on $[0,T)$.

We are going to show that $\overbar{Y}$ solves (\ref{eq:Y-ode-t}), which implies that $\overbar{Y}$ is the maximal positive solution to (\ref{eq:Y-ode-t}). It suffices to show $\overbar{Y}(T)=0$.
For any $j\ge1$, by $\overbar{\eta}\leq Y_j(0)$ and (\ref{ineq:m2C}), applying the comparison theorem to the forward ODEs satisfied by $\overbar{Y}$ and $Y_j$ yields that $\overbar{Y}\leq Y_j$ on $[0,\delta]$ for any $\delta\in(0,T)$. Then $\overbar{Y}(T)\leq Y_j(T)=\varepsilon_j\rightarrow0$ as $j\rightarrow\infty$. Thus, $\overbar{Y}(T)=0$. 
\end{proof}

\begin{rem}\label{rem:approximate eta}
From the proof of the previous proposition, we know that
\begin{equation*} 
Y_j(0)\searrow\overbar{\eta}=\overbar{Y}(0)=\eta^*\quad\text{ as } j\nearrow\infty,
\end{equation*} 
where $\overbar{Y}$ is the maximal positive solution to ODE (\ref{eq:Y-ode-t}).
This provides an approximating sequence $\{Y_j(0),j\ge1\}]$ of $\eta^*$ and an approximating sequence  $\{Y_j,j\ge1\}$ of the maximal positive solution $\overbar{Y}$. The approximation is implementable since ODE \eqref{eq:Y-ode-t-epj} admits a unique solution.  
\end{rem}

\begin{prop}
\label{prop:exist-Y-t-m} Let $\gamma<0$ and $\bm{\mu}\ne0$. Under Assumptions \ref{assu:DistFun-t}, \ref{assu:h-t} and \ref{ass:zeta}, $Y$ is a positive solution to the backward ODE (\ref{eq:Y-ode-t}) if and only if $Y$ solves the forward ODE (\ref{eq:Y-ode-t-forward}) for some $\eta\in(0,\eta^*]$.
\end{prop}

\begin{proof} Based on Proposition \ref{prop:Y_M}, the 
result follows from Lemmas \ref{lem:multiple-1}--\ref{lem:multiple-2}.
\end{proof}

\subsection{Uniqueness of the Positive Solution}

\begin{prop}
\label{prop:exist-Y-t-u} 
Let $\gamma<0$ and $\bm{\mu}\ne0$. Under Assumptions \ref{assu:DistFun-t}, \ref{assu:h-t}(i,ii) and \ref{ass:zeta}(ii), 
if condition \eqref{eq:cond:h:delta} holds, then ODE (\ref{eq:Y-ode-t}) admits a unique solution, which is positive. 
\end{prop}

\begin{proof}
Let $\delta\in(0,T)$ be given as in condition \eqref{eq:cond:h:delta}.
\begin{description}
\item{(I)} We show that, for sufficiently small $\varepsilon\in(0,\delta)$, the ODE
\begin{equation}
\begin{cases}
Y'(t)=-\theta^{2}m^{2}\left(t,\sqrt{Y(t)}\right), & t\in[T-\varepsilon,T),\\
Y(T)=0
\end{cases}\label{eq:Y-ode-epsilon}
\end{equation}
admits a unique solution and the solution is positive. To this end, let $Z(t)=\sqrt{-Y'(t)}$ and transform ODE \eqref{eq:Y-ode-epsilon} into the following integral equation
\begin{equation}
Z(t)=\theta m\left(t,\sqrt{\int_{t}^{T}Z^{2}(s)\dt s}\right),\quad t\in[T-\varepsilon,T).\label{eq:z-int}
\end{equation}

\item[(I.1)] We first show that integral equation \eqref{eq:z-int} admits a unique solution if $\varepsilon\in(0,\delta)$ is  sufficiently small. 

Since $0\le m(t,x)\leq1$ for all $(t,x)\in[0,T)\times[0,\infty)$, we know that $0\leq Z\leq\theta$ for every solution $Z$ to equation (\ref{eq:z-int}). For any $0\leq s<t<T$, let
$$\mathbb{Z}(s,t)=\left\{ Z\in L^{\infty}(s,t)\,:\,0\leq Z\leq\theta\right\} .$$
Obviously, $\mathbb{Z}(s,t)$ is a closed subset of $L^{\infty}(s,t)$.

By $0<m(t,x)\leq1$ and $h\left(t,-\gamma x\right)\geq1$ for all $(t,x)\in[T-\varepsilon,T)\times(0,\infty)$
and by \eqref{eq:m_x}, we have
\begin{equation}
\begin{split}
\left|m_{x}(t,x)\right|\leq\left|\frac{\gamma xh_{xx}\left(t,-\gamma x\right)+h_{x}\left(t,-\gamma x\right)}{x^{2}}\right|+\left|\gamma x\left[\frac{h_{x}\left(t,-\gamma x\right)}{x}\right]^{2}\right|,&\\
(t,x)\in[T-\varepsilon,T)\times(0,\infty).
\end{split}
\label{eq:|m_x|}
\end{equation}

For any $(t,x)\in[T-\varepsilon)\times(0,\theta^2\delta]$, it follows from the mean value theorem that there exists $y_i\in(0,-\gamma x)$, $i=1,2$, such that 
\begin{align*}
h_{x}\left(t,-\gamma x\right)=&-\gamma xh_{xx}(t,0)+\frac{1}{2}\gamma^2x^2h_{xxx}(t,y_1),\\
h_{xx}(t,-\gamma x)=&h_{xx}(t,0)-\gamma xh_{xxx}(t,y_2),
\end{align*}
which imply
\begin{align*}
\left|\frac{\gamma xh_{xx}\left(t,-\gamma x\right)+h_{x}\left(t,-\gamma x\right)}{x^{2}}\right|=&\gamma^2\left|\frac{1}{2}h_{xxx}(t,y_1)-h_{xxx}(t,y_2)\right|,\\
\left|\frac{h_{x}\left(t,-\gamma x\right)}{x}\right|=&\left|-\gamma h_{xx}(t,0)+\frac{1}{2}\gamma^2xh_{xxx}(t,y_1)\right|.
\end{align*}

Therefore, by condition \eqref{eq:cond:h:delta}, there exists a constant
$L>0$ such that $\left|m_{x}(t,x)\right|\leq L$ for all $(t,x)\in[T-\varepsilon,T)\times(0,\theta^{2}\delta]$. 

Consider the operator $\mathcal{T}:\mathbb{Z}(T-\varepsilon,T)\rightarrow\mathbb{Z}(T-\varepsilon,T)$
defined by 
\[
\left(\mathcal{T}Z\right)(t)\triangleq\theta m\left(t,\sqrt{\int_{t}^{T}Z^{2}(s)\dt s}\right),\quad t\in[T-\varepsilon,T).
\]
For any $Z_{i}\in\mathbb{Z}(T-\varepsilon,T)$, $i=1,2$, we have
\begin{align*}
\left\Vert \Tc Z_{1}-\Tc Z_{2}\right\Vert _{\infty}  \leq & \theta L\sup_{t\in[T-\varepsilon,T)}\left|\sqrt{\int_{t}^{T}Z_{1}^{2}(s)\dt s}-\sqrt{\int_{t}^{T}Z_{2}^{2}(s)\dt s}\right|\\
  \leq & \theta L\sup_{t\in[T-\varepsilon,T)}\sqrt{\int_{t}^{T}\left|Z_{1}(s)-Z_{2}(s)\right|^{2}\dt s}\\
  \leq & \theta L\sqrt{\varepsilon}\left\Vert Z_{1}-Z_{2}\right\Vert _{\infty}.
\end{align*}
Choosing $\varepsilon<\min\{1/(\theta^{2}L^{2}),\delta\},$ we have that
$\mathcal{T}$ is a contraction on $\mathbb{Z}(T-\varepsilon,T)$.
Thus, there exits a unique solution $Z$ to (\ref{eq:z-int}). 

\item[(I.2)] Let $Z$ be the solution obtained in (I.1). Let $Y(t)=\int_t^TZ^2(s)\dt s$, $t\in[T-\varepsilon,T)$. Then $Y$ is the unique solution to ODE \eqref{eq:Y-ode-epsilon}. We are going to show  $Y$ is positive. 

Since $m(t,0)=1/(1-\gamma h_{xx}(t,0))>0$ for all $t\in[T-\varepsilon,T)$ and  $m(t,x)>0$  for all $(t,x)\in[T-\varepsilon,T)\times(0,\infty)$,
we have $Z(t)>0$ for all $t\in[T-\varepsilon,T)$.
Then $Y(t)=\int_{t}^{T}Z^{2}(s)\dt s>0$ for all $t\in[T-\varepsilon,T)$. 

\item{(II)} We show the existence and uniqueness of the solution to the following ODE:
\begin{equation}
\begin{cases}
Y'(t)=-\theta^{2}m^{2}\left(t,\sqrt{Y(t)}\right), & t\in[0,T-\varepsilon),\\
Y(T-\varepsilon)=\beta,
\end{cases}\label{eq:Y-ode-epsilon-beta}
\end{equation}
where $\varepsilon\in(0,\delta)$ and $\beta\in(0,\theta^2\varepsilon]$.

Similar to (\ref{ineq:m2C}), there exits
a constant $C>0$ such that 
\begin{eqnarray*}
\left|\frac{dm^{2}(t,\sqrt{x})}{dx}\right|<C\quad\forall(t,x)\in[0,T-\varepsilon]\times[\beta,\theta^{2}T].
\end{eqnarray*}

Hence, applying the fixed point theorem, one know that ODE \eqref{eq:Y-ode-epsilon-beta}
has a unique solution, which satisfies
$\beta\leq Y(t)\leq\theta^{2}T$, $t\in[0,T-\varepsilon]$.  
\end{description}
A combination of (I) and (II) yields the conclusion of the proposition.
\end{proof}

\begin{rem} Part (I.1) of the previous  proof is similar to the proof of  
\citet*[Theorem 3.5]{lwxy24}. It is provided here for the readers’ convenience.  
\end{rem}

\section{ODE \eqref{eq:Y-ode-log-t}}\label{app:ode-log}
The proofs of the following propositions are either trivial or similar to those in Appendix \ref{sec:ode-t}.
\begin{prop}
\label{prop:exist-Y-t-log-0} In addition to Assumption \ref{assu:DistFun-t}, assume that  $h_x(\cdot,0)$ is continuous on $[0,T)$ and $\inf_{t\in[0,T)}h_{x}(t,0)>0$, then $0$ is the unique solution to ODE (\ref{eq:Y-ode-log-t}). 
\end{prop}

\begin{prop}\label{prop:exist-Y-t-log}
Let $\bm{\mu}\ne0$.  In addition to Assumption \ref{assu:DistFun-t}, assume that  $h_x(\cdot,0)$ is continuous on $[0,T)$.
\begin{description}
\item[(i)] If $h_{x}(\cdot,0)\geq0$ on $[0,T)$ and $\limsup_{t\uparrow T}\frac{h_{x}(t,0)}{\theta\sqrt{T-t}}<1$, then ODE (\ref{eq:Y-ode-log-t}) has a positive solution;
\item[(ii)] If $h_{x}(\cdot,0)>0$ on $[0,T)$ and $\limsup_{t\uparrow T}\frac{h_{x}(t,0)}{\theta\sqrt{T-t}}<1$, then the backward ODE (\ref{eq:Y-ode-log-t}) has a maximal positive solution; In this case, $Y$ is a positive solution to the backward ODE (\ref{eq:Y-ode-log-t}) if and only if $Y$ solves the forward ODE \eqref{eq:Y-ode-log-t-forward}
for some $\eta\in(0,\eta^*]$, where
\[
\eta^*=\sup\{Y(0):Y\text{ is a positive solution to (\ref{eq:Y-ode-log-t})}\};
\]
\item[(iii)] If $h_{x}(\cdot,0)\geq0$ on $[0,T)$ and there exists $\delta\in(0,T)$
such that $h_{x}(t,0)=0$ for all $t\in[T-\delta,T)$, then ODE (\ref{eq:Y-ode-log-t}) admits a unique solution, which is positive. 
\end{description}
\end{prop}

\end{appendices}

\bibliographystyle{abbrvnat} 
\bibliography{mybib}
\end{document}